\documentclass[prl,10pt,twocolumn,aps]{revtex4-1}

\usepackage[utf8]{inputenc}
\usepackage[T1]{fontenc}
\usepackage{amsmath}
\usepackage{amssymb}
\usepackage{graphicx}
\usepackage[colorlinks]{hyperref}
\usepackage{bm}
\usepackage{xcolor}
\usepackage{adjustbox}
\usepackage{soul}
\usepackage[normalem]{ulem}

\global\long\def\ket#1{\left| #1\right\rangle }
\global\long\def\bra#1{\left\langle #1 \right|}

\global\long\def\av#1{\left\langle #1 \right\rangle }

\global\long\def\Tr{\text{Tr}}

\global\long\def\bs#1{\boldsymbol{#1}}

\definecolor{applegreen}{rgb}{0.55, 0.71, 0.0}

\newcommand{\beginsupplement}{%
        \setcounter{table}{0}
        \renewcommand{\thetable}{S\arabic{table}}%
        \setcounter{figure}{0}
        \renewcommand{\thefigure}{S\arabic{figure}}
        \setcounter{section}{0}
        \renewcommand{\thesection}{S\arabic{section}}%
        \setcounter{section}{0}
        \renewcommand{\thesubsection}{S\arabic{section}.\arabic{subsection}}%
        \setcounter{equation}{0}
        \renewcommand{\theequation}{S\arabic{equation}}%
     }

\makeatother

\begin{document}

\title{ Voltage-Driven Breakdown of Electronic Order}

\author{Miguel M. Oliveira}
\email{miguel.m.oliveira@tecnico.ulisboa.pt}
\affiliation{CeFEMA, Instituto Superior T\'ecnico, Universidade de Lisboa Av. Rovisco Pais, 1049-001 Lisboa, Portugal}

\author{Pedro Ribeiro}
\email{ribeiro.pedro@tecnico.ulisboa.pt}
\affiliation{CeFEMA, Instituto Superior T\'ecnico, Universidade de Lisboa Av. Rovisco Pais, 1049-001 Lisboa, Portugal}
\affiliation{Beijing Computational Science Research Center, Beijing 100193, China}

\author{Stefan Kirchner}
\email{kirchner@nycu.edu.tw}
\affiliation{Department of Electrophysics, National Yang Ming Chiao Tung University, Hsinchu 30010, Taiwan}

\begin{abstract}
The non-thermal breakdown of a Mott insulator has been a topic  of great theoretical and experimental interest with technological relevance. 
Recent experiments have found a sharp non-equilibrium insulator-to-metal transition that is accompanied by hysteresis, a negative differential conductance and lattice deformations. However,  a thorough understanding of the underlying breakdown mechanism  is still lacking.
Here, we examine a scenario in which the breakdown is induced by chemical pressure in a paradigmatic model of interacting spinless fermions on a chain coupled to metallic reservoirs (leads).
For the Markovian regime, at infinite bias, we qualitatively reproduce several established results.
Beyond infinite bias, we find a rich phase diagram where the nature of the breakdown depends on the coupling strength as the bias voltage is tuned up, yielding different current-carrying non-equilibrium phases. 
For weak to intermediate coupling, we find a conducting CDW phase with a bias-dependent ordering wave vector. At large interaction strength, the breakdown connects the system to a charge-separated insulating phase. 
We find  instances of  hysteretic behavior,   sharp current onset and   negative differential conductance. 
Our results can help to shed light on recent experimental findings that address current-induced Mott breakdown. 
\end{abstract}

\maketitle

\section*{Introduction}

The current-voltage characteristics has been a central quantity in characterizing materials. 
In weakly correlated semi-conductors, a current ensues when the voltage drop surpasses the energy gap in the low-temperature limit. In correlated (i.e. Mott) insulators the situation is less clear. Here, voltages much smaller than the interaction-induced equilibrium gap could already disturb the electronic distribution function and destabilize the insulating state. 
Furthermore, the non-equilibrium electronic distribution function in the presence of interactions could stabilize intrinsic out-of-equilibrium phases. 
This makes the understanding of the interplay of non-equilibrium conditions and interaction effects in strong-correlated phases of fundamental interest in the exploration of novel phases of matter unrestricted by equilibrium constraints (e.g. fluctuation dissipation relations).  

To ensure that the electronic distribution function does not reach local equilibrium, samples should be clean enough for the transport in the current-carrying phases to be ballistic, with the impurity scattering mean-free path of the order of the sample size. At the same time, temperatures have to be low enough to allow for non-trivial correlated phases.  
Several experimental studies have recently been conducted in this transport regime for a number of correlated materials~\cite{Alexander99,Nakamura2013,Zhang2019,Bertinshaw2019,Fursich2019,Cirillo2019,Gauquelin2023,Curcio2023,Cao2020,Diener_2018,Rahaman2021,Maklar2021,Kanki2012,Yamanouchi99,Taguchi2000,Imada98}. 
The extent to which the various experimental studies were able to establish that non-thermal effects do induce the Mott breakdown remains at present unclear~\cite{Alexander99,Nakamura2013,Zhang2019,Bertinshaw2019,Fursich2019,Cirillo2019,Gauquelin2023,Curcio2023,Cao2020}.
While the nature of the transition is yet not fully understood, it has become clear that in most cases the electronic transition is accompanied by a first order lattice deformation upon applying voltage, implying the correlated phase involves both electronic and lattice degrees of freedom. This scenario is reminiscent of the certain equilibrium Mott transitions where lattice degrees of freedom also play an important role \cite{Park2013}.  

Two main frameworks have been invoked to explain the collapse of the Mott phase. 
The first is that of a thermal-like breakdown. 
Among the types of metal-insulator transitions in this class the non-equilibrium steady-state can be mapped into an local equilibrium state with thermalized electronic and lattice degrees of freedom~\cite{Diener_2018,Yamanouchi99,Taguchi2000,Frohkich37,Frohkich43}.  
However, such scenarios seem incompatible with some of the experimental work that has ruled out effects due to Joule heating~\cite{Zhang2019,Yamanouchi99,Taguchi2000}.
Recently, intrinsic non-equilibrium mechanisms of the energy transfer between the lattice and the electronic degrees of freedom in the presence of a strong electric field have been studied~\cite{Han2018, Han2023, Chen2024, Chiriaco2018, Picano2023} which also rely on effective temperature effects to trigger the transition. 

The second framework is Landau-Zener breakdown where the non-equilibrium drive is a constant electric field applied across the sample. Through Peierls substitution, this effect can be modeled in  a system with periodic boundary conditions pierced by a  linearly increasing  time-dependent flux . For small flux, the effective time-dependent Hamiltonian changes slowly when compared with the Mott gap,  and thus, the system remains in its groundstate. Eventually, as the flux increases,  transitions to excited states start to occur.
 As there is no need for an explicit treatment of the reservoirs the problem is amenable to various standard methods like diagonalization~\cite{Oka2003}, the density-matrix renormalization group (DMRG)~\cite{Oka2005,Meisner2010}, dynamical mean-field theory (DMFT)~\cite{Aron2012,Eckstein2010,Eckstein2011,Li2015,Neumayer2015,Lee14}, and perturbative methods~\cite{Lenarcic2012,Mierzejewski2011}.
These studies revealed a qualitative scenario that can be interpreted as the many-body analog of the Landau-Zener mechanism observed in band insulators~\cite{Oka2003} where a characteristic energy scale sets a threshold above which a field-induced metallic phase forms. As it turned out, this breakdown energy scale is typically overestimated  as compared to the experimentally measured values. 

A third scenario for the non-thermal destruction of a correlated insulator, that has been overlooked, is through the effect of chemical pressure induced by the thermodynamic inbalance in the electronic distribution functions of the leads. For a ballistic metal,  the screening length is of the order of the lattice spacing and electric field gradients are negligible in the steady-state regime. Thus, approaching the breakdown from the metallic phase, considering a thermodynamic inbalance-dominated regime is a natural choice. 
In the insulating state, on the other hand, considerable corrections are expected, although the chemical pressure effect might still be dominant near the breakdown bias.

Previous works addressing this transport regime have focused on the Hubbard model in one and higher dimensions at the mean-field level~\cite{Ribeiro2016,Dutta2020} and found  different behaviors as a function of bias and interaction strength. In one dimension, the transition to the metallic state occurs  in the large bias regime through a bias-induced patterned phase where the current is transported through in-gap states~\cite{Ribeiro2016}. 
In two dimensions, one finds a growing metallic region near the leads with a characteristic length-scale that diverges at the transition~\cite{Dutta2020}.  
Recently, using an effective Langevin dynamics~\cite{Dutta2022}, it was shown that the 3d Hubbard model supports a first order transition with the bias where an hysteretic regime is stabilized which is absent in the 2d case. 
These works show that thermodynamic inbalance effects can induce Mott-breakdown and thereby stabilize non-equilibrium metallic phases.

Motivated by the experimental situation and by the debate around the various scenarios for Mott breakdown, we focus on a paradigmatic model, schematically illustrated in Fig.~\ref{fig:pd}-(a), featuring a bias-induced CDW-to-metal transition and assuming that the relevant mechanism is thermodynamic inbalance. We consider the zero temperature limit where the interplay between non-equilibrium conditions and correlations is expected to be most pronounced. For simplicity, we consider a purely electronic system with no lattice degrees of freedom.  This enables us to benchmark our results in various limits.
For this model, we compute the mean-field phase diagram as a function of the bias and interaction strength and study the response functions of various observables at the RPA level.
We find that already an half-filled tight-binding fermionic chain with nearest-neighbour density-density interactions, $U$, under bias, $V$, possesses an extremely rich phase diagram, see Fig.~\ref{fig:pd}-(b). Besides the equilibrium  conducting and CDW phases with wavevector $\pi$, we unveil a CDW with a bias-dependent wavevector at small $V$ and intermediate $U$, and several conducting, strongly-interacting, disordered phases.

A large body of theoretical work addresses boundary driven Markovian systems, including exact solutions~\cite{Marko2,Marko3,Prosen3,Prosen4,Prosen6,Prosen7} and DMRG results~\cite{Prosen1,Benenti1,Benenti2,Marko1,Prosen2,Prosen5,Marko4,Marko5,Mendoza1,Mendoza2,landi2021nonequilibrium}. In the present setup, we demonstrate that the Markovian regime can be obtained in the large bias limit. Thus, conceptually, we link these well-known results to the experimentally relevant finite bias regime.

The paper is organized as follows. We first introduce the model and the details of the mean-field non-equilibrium methods. In the Results section, we discuss our findings starting with the week-coupling regime. For strong coupling, we provide a simplified model, that provides an intuitive physical picture for the mean-field predicted phases. We also discuss the large bias regime which encompasses the Markovian limit. In the final section, we discuss implications of our findings.

\section*{Model and Methods}

\subsection*{Non-equilibrium Fermionic Chain}

We consider a chain of spinless fermions with nearest-neighbor interactions coupled at its edges to metallic reservoirs which are held at their respective chemical potentials. A sketch of this model is provided in Fig.\ref{fig:pd}-(a). 
The Hamiltonian of the full system is given by $H= H_\text{C} + \sum_l H_l + \sum_l H_{\text{C},l}$, where $H_\text{C}$ is the Hamiltonian of a chain of $L$ sites. 
The reservoir's label, $l={1,L}$, indicates the site at which it couples to the chain, $H_l$ denotes the reservoir Hamiltonian at site $l$ and $H_{\text{C},l}$ is the chain-reservoir coupling.  We chose 
\begin{align}
H_\text{C} &= -t \sum_{r=1}^{L-1} \left( c^\dagger_r \, c_{r+1} + c^\dagger_{r+1} \, c_r \right) \nonumber \\
&\quad + U \sum_{r=1}^{L-1} \left( c^\dagger_r \, c_r - \frac{1}{2} \right) \left( c^\dagger_{r+1} \, c_{r+1} - \frac{1}{2} \right) \quad ,
\label{model_ham}
\end{align}
where $t$ is the nearest-neighbor hopping integral and we choose units of energy such that $t=1$. $U$ is the strength of the repulsive interaction between nearest-neighbors. $c^\dagger_r$ ($c_r$) is the creation (annihilation) operator for a spinless fermion at site $r$. 
We consider metallic featureless leads with a bandwidth much larger than any of the system's energy scales. 
In this case, the system-reservoir coupling is entirely specified by an hybridization energy $\Gamma_l$, corresponding to the rate of escape of an electron in the site adjacent to the lead. The chemical potential of a reservoir is denoted by $\mu_l$. 
Below we provide details of the system-reservoir coupling. 
Our primary interest is the generalization of the equilibrium zero temperature phase diagram once a bias voltage is applied across the metallic leads.

We have written Eq.~(\ref{model_ham}) in a manifestly particle-hole symmetric form to access the regime where the equilibrium state is ordered. We also set the average chemical potential to zero, i.e., $(\mu_1+\mu_L)/2=0$, and define the bias via $V=\mu_1-\mu_L$. Thus, $V=0$ corresponds to a half-filled chain. For $V\neq 0$, the system is still invariant under the combination of a particle-hole transformation and a mirror symmetry around the center of the chain.

The model thus defined is a paradigmatic model for which some limits are known. 
In particular, the equilibrium groundstate of the isolated chain is exactly known~\cite{FranchiniBook,Zotos,Benz}. 
It is recovered in the present model for $V=0$ only sufficiently far from the leads, beyond a characteristic length-scale that becomes arbitrarily small in the limit $\Gamma_l\to 0$. 
The equilibrium quantum phase transition between a Luttinger liquid and an ordered charge density-wave (CDW) occurs at $U=2$.

In the $V\to\infty$ limit, the left (right) reservoir can only give (receive) particles~\cite{Ribeiro2015}. In this limit, the model can be mapped into a boundary-driven Markovian chain for which the non-equilibrium steady-state is exactly known~\cite{Prosen3,Prosen4}. 
It features a transition between a ballistic transport regime and a phase separated insulating state at $U=2$.

For $U=0$, the model turns into a non-interacting conducting tight-binding chain of bandwidth $W=4t$~\cite{Ribeiro2017}. Thus, for a bias smaller than $W$, the system is in a non-saturated conducting regime, where the differential conductivity does not vanish. For $V>W$, a saturated current regime ensues.

\subsection*{System-leads coupling}
We assume metallic reservoirs with an Hamiltonian given by
\begin{equation}
H_l = \sum_k \epsilon_{l,k} \, f^\dagger_{l,k} \, f_{l,k} \quad ,
\label{Ham_reservoir}
\end{equation}
corresponding respectively to the left ($l=1$) and right ($l=L$) reservoir. $\epsilon_{l,k}$ is the energy of the $k$'th mode of the $l$'th reservoir, with corresponding creation and annihilation operators $f^\dagger_{l,k}$ and $f_{l,k}$. The chain-reservoir coupling to the $l$-th reservoir is given by
\begin{equation}
H_{\text{C},l} = \frac{J_l}{\sqrt{N}} \sum_k \left( c^\dagger_l \, f_{l,k} + f^\dagger_{l,k} \, c_l \right) \quad ,
\label{Ham_res_chain}
\end{equation}
where $J_l$ is the hopping integral and $N$ the total number of modes.

The bandwidth of the otherwise featureless leads are taken to be much larger than the system's energy scales. In this case, the only quantity that characterizes the leads is their chemical potential, $\mu_l$, and 
hybridization constant $\Gamma_l = 2\pi \, J^2_l \, \rho_l$, where $\rho_l$ is the density of states of lead $l$. In the supplemental material (SM), see Ref.~\cite{SupMat}-S1, we provide a detailed description of the wide-band limit. 
In this limit, the coupling with the leads induce a self-energy contribution to the system's Green function whose retarded, advanced and Keldysh components are given, respectively by
\begin{align}
& \bm{\Sigma}^{R/A} (t,t') = \mp i \delta(t-t') \sum_l \frac{\bm{\Gamma}_l}{2} \nonumber \\
& \bm{\Sigma}^K (t,t') = -i \sum_l \bm{\Gamma}_l \int \frac{d\omega}{2\pi} \tanh\left[ \frac{\beta_l(\omega-\mu_l)}{2} \right] e^{-i \omega(t-t')},
\label{self_energies}
\end{align}
where $\beta_l$ and $\mu_l$ are the inverse temperature and chemical potential of the $l$'th reservoir; and $\bm{\Gamma}_l$ are matrices in the position basis such that 
$\bm\Gamma_l(r,r')= \Gamma_l \, \delta_{r,r'}\, \delta_{r,l}$ (see SM~\cite{SupMat}-S1). 
In all  numerical results presented here, we set $J_l=\rho_l=1$ for each reservoir. Our analysis is conducted at zero temperature, where the hyperbolic tangent gets reduced to a sign function.

\subsection*{Mean-field Order Parameter}

Within the non-equilibrium mean-field approach, we decouple the interaction via a Hubbard-Stratonovich (HS) transformation $U \,c^\dagger_r \, c_r \, c^\dagger_{r+1} \, c_{r+1} \to -\frac{1}{U} \, \varphi_r \, \varphi_{r+1} + \varphi_r \, c^\dagger_r \, c_r$ , where $\varphi_r$ is a bosonic real field. The fermionic fields can then be formally integrated out in the path integral.  

After performing the Keldysh rotation the action can be written in terms of the so-called classical and quantum components of the HS field $\left( \varphi^c_r, \varphi^q_r \right)$. In these variables, minimizing the action leads to the self-consistent saddle-point equations
\begin{align}
\varphi^q_r(t) &= 0 \nonumber \\
%\varphi^c_r(t) &= -i \, \frac{U}{2} \left[ G^K_{r-1,r}(t,t) + G^K_{r,r+1}(t,t) \right] \\
\varphi^c_r(t) &= \, \frac{U}{2} \left[ \langle n_{r-1}(t)\rangle + \langle n_{r+1}(t) \rangle \right]
\quad ,
\end{align}
where $n_0(t)=n_{L+1}(t)=0$. 
In the steady-state regime, we drop the time dependence, $\varphi^c(t)=\varphi^c$. 

Here, it is useful to introduce a mean-field single-particle non-hermitian Hamiltonian $\bm{K}= \bm{H}_C + \bm{\varphi}^c - \frac{i}{2} \sum_l \bm{\Gamma}_l$, where $\left[ \bm{\varphi}^c \right]_{r,r'} = \varphi^c_r \, \delta_{r,r'}$ and $\left[\bm{H}_C \right]_{r,r'} = -t (\delta_{r,r'+1}+\delta_{r+1,r'}) - \delta_{r,r'} U_r $ with $U_r= U/2$ for $r=1,L$ and $U_r=U$ otherwise.

We focus on the steady-state regime obtained for asymptotically large times after turning on the system-reservoir coupling. In this regime, time-translation invariance allows us to write the different components of the Green function in frequency space as  
\begin{align}
\bm{G}^R(\omega) &= \left[\bm{G}^A(\omega) \right]^\dagger = \left( \omega - \bm{K} \right)^{-1} \nonumber \\
%\bm{G}^A(\omega) &= \left[\bm{G}^R(\omega) \right]^\dagger \nonumber \\
\bm{G}^K(\omega) &= \bm{G}^R(\omega) \, \bm{\Sigma}^K (\omega) \, \bm{G}^A(\omega) \quad ,
\label{Green_freq}
\end{align}
where $\omega$ is the frequency and the labels $R$, $A$ and $K$ refer respectively to the retarded, advanced and Keldysh components of the Green function (see also SM~\cite{SupMat}-S1). 
From the Green functions, we calculate the different observables. Of particular interest are the single-particle equal-time correlators, conveniently combined into the so-called covariance matrix $\varrho_{r,r'} =\av{c_r^\dagger c_{r'}} $ that can be expressed in terms of $\bs G^K$ as
\begin{equation}
\bm{\varrho} = \frac{1}{2} \left[ \bm{1} -i \int \frac{d\omega}{2\pi} [\bm{G}^K(\omega)]^T \right] \quad.
\end{equation} 
The frequency integration is explicitly done in SM~\cite{SupMat}-S2.
From the covariance matrix we obtain the occupation numbers $n_r= \langle c^\dagger_r \, c_r \rangle = \varrho_{r,r}$ and the electron current $J_r = -i t \left\langle  c^\dagger_{r} \, c_{r+1} - c^\dagger_{r+1} \, c_{r}  \right\rangle = -i t \left( \varrho_{r,r+1} - \varrho_{r+1,r} \right)$.

\subsection*{RPA Susceptibility}

The form of  the charge susceptibility consistent with the  non-equilibrium mean-field approximation is given in terms of a random phase approximation (RPA) on the Keldysh contour,
\begin{align}
& \bm{\chi}_{\text{RPA}}  = \bm{\chi}_0  \left[ \bm{1} + \bm{U}  \, \bm{\chi}_0  \right]^{-1} \quad , 
\end{align} 
where $\bm{U}_{r,r'}(z,z')= \delta(z-z') (\delta_{r,r'+1}+ \delta_{r,r'-1} ) U/2$. 
Its retarded component corresponds to 
\begin{align}
\bm{\chi}_\text{RPA}^R (\omega) = \bm{\chi}_0^R(\omega) \left[ \bm{1} + \bm{U} \, \bm{\chi}_0^R(\omega) \right]^{-1} \quad ,
\label{ChiR_RPA}
\end{align} 
where the bare retarded susceptibility is given by
\begin{align}
& \left[ \bm{\chi}_0^R (\omega) \right]_{r,r'} = \nonumber  \\
&  \frac{i}{2} \int \frac{d\nu}{2\pi} \left[ G^R_{r,r'}(\nu) \, G^K_{r',r}(\nu-\omega) + G^K_{r,r'}(\nu) \, G^A_{r',r}(\nu-\omega)  \right] .
\label{susceptibility}
\end{align}

It is also useful to define the Keldysh component of the bare susceptibility
\begin{align}
& \left[ \bm{\chi}_0^K (\omega) \right]_{r,r'} = \nonumber \\
&  \frac{i}{2} \int \frac{d\nu}{2\pi} \left[ G^R_{r,r'}(\nu) \, G^A_{r',r}(\nu-\omega) + G^A_{r,r'}(\nu) \, G^R_{r',r}(\nu-\omega) \right] \nonumber \\ 
& + \frac{i}{2} \int \frac{d\nu}{2\pi} \, G^K_{r,r'}(\nu) \, G^K_{r',r}(\nu-\omega)    \quad .
\end{align}

\section*{Results}

%%%%%%%%%%%%%%%%%%%%%%% Figure 1 %%%%%%%%%%%%%%%%%%%%%%%%%%
\begin{figure}[t!]
\centering
\includegraphics[width= 0.5 \textwidth]{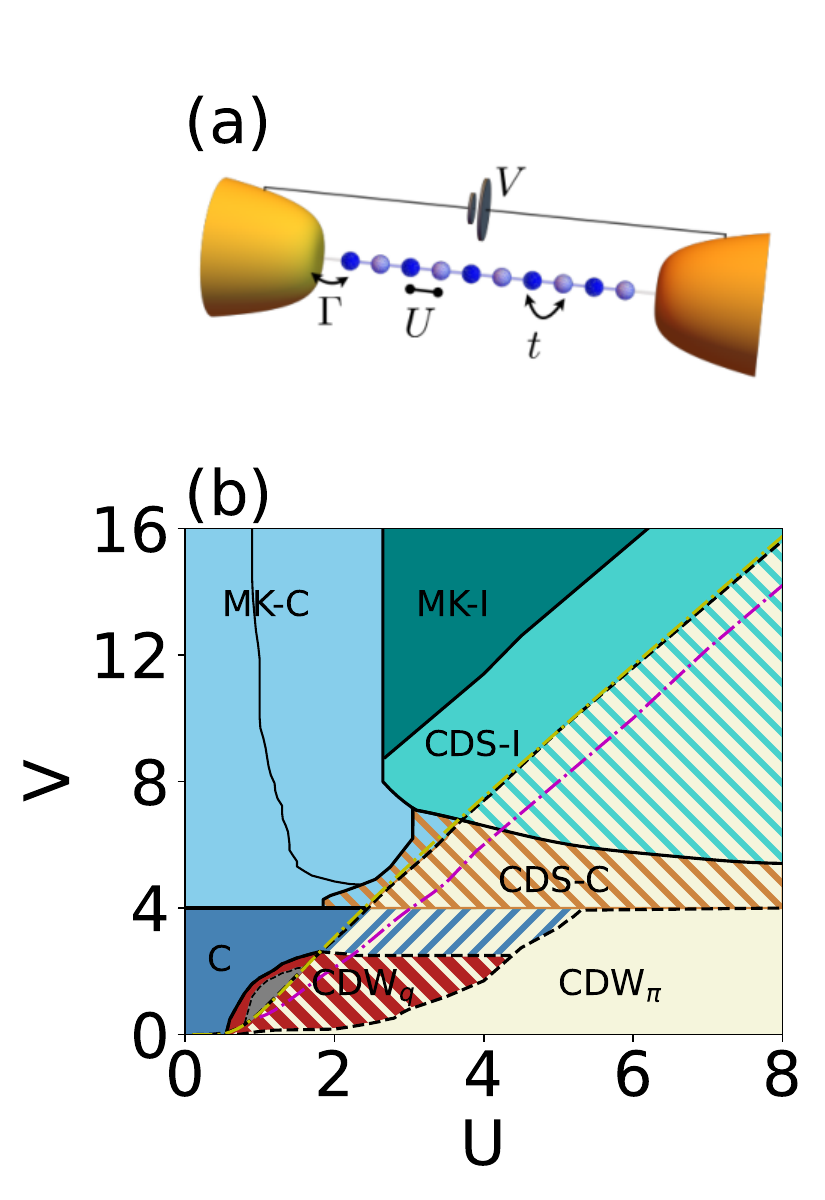} 
\caption{(a) Sketch of the boundary-driven fermion chain model. (b) Mean-field phase diagram showcasing the non-equilibrium phases: charge density waves with $q=\pi$ ($\text{CDW}_\pi$) and $V$-dependent $q$ ($\text{CDW}_q$), a non-saturated conducting phase (C), a ballistic regime with a saturated current (MK-C), an insulating charge separated phase (MK-I), and two other charge separated phases, one insulating (CDS-I) and one conducting  (CDS-C).  
The gray region corresponds to an unstable regime, highly sensitive to initial conditions. Continuous and dashed lines mark second-order and first-order phase transition respectively. 
Striped regions signal metastability of several mean-field solutions and are colored accordingly. The dash-dotted yellow line marks the equilibrium gap, which coincides with the end of $\text{CDW}_\pi$. The dash-dotted purple line marks the appearance of boundary defects. 
}
\label{fig:pd}
\end{figure}
%%%%%%%%%%%%%%%%%%%%%%%%%%%%%%%%%%%%%%%%%%%%%%%%%%%%%%%%%%

The main result of this work is the mean-field phase diagram shown in Fig.~\ref{fig:pd}-(b). This phase diagram features several phases separated by non-equilibrium quantum phase transitions. 
Besides the conducting (C) phase and the charge density ordered state of wave-vector $q=\pi$ ($\text{CDW}_\pi$)  that connect with the respective equilibrium phases, we find a charge density wave characterized by a $V$-dependent wave-vector ($\text{CDW}_q$). 
At larger voltages, we find a saturated conducting phase (MK-C), phase separated regimes that can be conducting (CDS-C) or insulating (CDS-I) and finally, a Markovian insulating state (MK-I). We also established that MK-C and MK-I connect to the respective states found in the Markovian limit for $V\to\infty$. 
Inside the MK-C, an additional transition line is found, terminating in a critical point (outside the plotted range, see SM~\cite{SupMat}-S6. 
In Fig.~\ref{fig:pd}-(b),  continuous and dashed lines separating these phases mark second-order and first-order phase transitions, respectively. An in-depth discussion of the various non-equilibrium phases and their properties is provided below.

For finite system sizes, the $V\rightarrow 0$ equilibrium phase transition is qualitatively recovered. 
In the thermodynamic limit, the mean-field incorrectly predicts an instability to a CDW to occur at $U=0$ while, for finite systems, this transition is pushed to non-zero $U$.
The drawbacks of the mean-field approximation for 1d equilibrium conditions, in particular its failure to capture the Luttinger liquid features of the gapless phase, are well known. However, when the available phase space increases, {\itshape e.g.}, in higher dimensions,  its predicting power also increases.  A likewise improvement is expected to occur at finite bias voltage where the increased range for energy fluctuations leads to better mean-field results. This expectation is borne out by our results at $V\to\infty$ where we find quantitative agreement between the mean-field and the exact solution.

\subsection*{Voltage driven CDW melting at weak coupling}

%%%%%%%%%%%%%%%%%%%%%%% Figure 1 %%%%%%%%%%%%%%%%%%%%%%%%%%
\begin{figure*}[t!]
\centering
\includegraphics[width= 1.0 \textwidth]{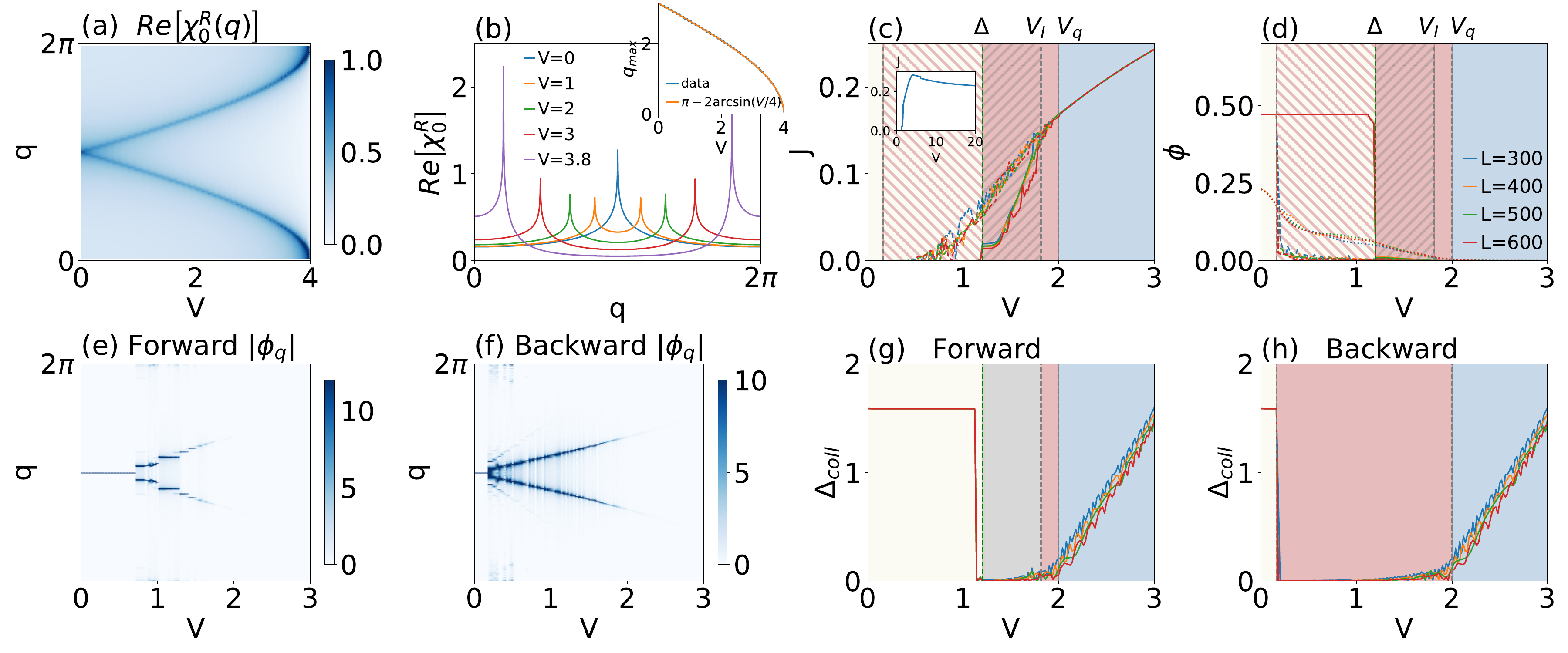} 
\caption{ Weak coupling. (a) Real part of the non-interacting susceptibility $\chi_0(\omega=0,q)$ as a function of the bias $V$ and the momentum $q$ computed for $L=1000$. (b) Cuts of (a) for different values of $V$. The inset compares the maximum value with the expression $q_\text{max}(V) = \pi \pm 2\arcsin(V/4t)$. 
(c) , (d) Current, $J$, and order parameters, $\phi_{\pi,q}$ , as a function of $V$ for different system sizes. $\phi_\pi$ is depicted for the forward (backwards) evolution with continuous (dashed) lines; the dotted lines depict the backwards $\phi_q$. Vertical dashed lines mark transitions between different phases: $\Delta$ (equilibrium gap that marks end of the CDW$_\pi$), $V_{I}$ (end of the intermediate region), and $V_q$ (end of the $\text{CDW}_q$). Intervals of $V$ corresponding to the different phases in Fig.\ref{fig:pd}-(b) are colored accordingly. Inset of (c): $J$ vs $V$ for a large voltage window.   
(e), (f) fourier transforms of the real space electron density vs $V$ for forward and backward evolution respectively, for $L=500$.
(g), (h) collective excitation gap $\Delta_\text{coll}$ as a function of $V$ for forward and backward evolution respectively. To suppress boundary effects the permittivity matrix is truncated, see SM~\cite{SupMat}-S5 for more details. Panels (c), (d), (g) and (h) follow the color code shown in (d). (c), (d) and (f)-(h) were computed for $U=1.2$ and (e) for $U=1.0$.
}
\label{fig:small_U}
\end{figure*}
%%%%%%%%%%%%%%%%%%%%%%%%%%%%%%%%%%%%%%%%%%%%%%%%%%%%%%%%%%

The existence of novel phases away from equilibrium can be inferred within the RPA through an analysis of the retarded susceptibility. 
Fig.~\ref{fig:small_U}-(b) shows the non-interacting retarded susceptibility $\chi^R_0(\omega=0,q)$ for different values of $V$. 
We observe that $q_\text{max}$, defined as the wave-vector maximizing $\chi^R_0$, varies with $V$ in a simple fashion $q_\text{max} = \pi - 2\arcsin(V/W)$, see the inset of Fig.~\ref{fig:small_U}-(b). 
Similarly to $V=0$,  $\chi^R_0(\omega=0,q_\text{max})$ diverges logarithmically with $L$  for any  $V<4$, see SM~\cite{SupMat}-S3.  Thus, imposing the density profile obtained for $U=0$, an RPA analysis based on Eq.~\eqref{ChiR_RPA} predicts the formation of an ordered state for infinitesimal  $U$.  For $V>W$, i.e., voltages larger than the bandwidth, $\chi^R_0(\omega=0,q)$  is smooth and does not diverge in the thermodynamic limit.

In the present case, the finite-scaling analysis is more delicate than for a closed systems with periodic boundary conditions, as the self-consistent parameter $\varphi_r$ acquires an inhomogeneous finite value for $U\neq 0$ due to the presence of the leads, even in the absence of long-range order. 
Nevertheless, the qualitative argument based on the divergence of the RPA susceptibility, obtained by imposing the order parameter to vanish, captures the novel bias-induced charge ordered phase, $\text{CDW}_q$ for finite $V$. The $\text{CDW}_q$ phase with order parameter $\phi_q = \frac{1}{L} \sum_r e^{i q r} \phi_r$ is depicted in red in Fig.~\ref{fig:pd}-(b).

In Fig.~\ref{fig:small_U}-(a), we show the Fourier transform of the retarded susceptibility $\chi^R_0(\omega=0,q)$ as a function of $V$. In (c) and (d)  the current, $J$, and the $\text{CDW}_\pi$ order parameter, $\phi_\pi$, are shown as a function of $V$ at $U=1.2$ for the forward (solid) and backward (dashed) voltage sweep of the hysteretic cycle. The backward evolution of $\phi_q$ is also shown as a dotted line. In the forward evolution, this phase is difficult to stabilize (see below). 

In equilibrium ($V=0$), $\text{CDW}_\pi$ has a finite energy gap $\Delta(U)$ (yellow dashed-dotted line on Fig.~\ref{fig:pd}-(b)). 
Starting from equilibrium and increasing $V$ adiabatically, we find that the $\text{CDW}_\pi$ is robust until $V=\Delta$. 
We model an adiabatic change in $V$ by solving the mean-field equations for $V+\delta V$, starting with the converged solution for $V$ as initial condition. 
The bias-induced transitions reported in Fig.~\ref{fig:pd}-(b) were obtained using this adiabatic procedure. 
Reminiscent of what occurs in equilibrium, the discontinuous nature of these transitions is accompanied by metastability and hysteresis. 

For a representative case of weak coupling ($U=1.2$), starting from some value of $V>W$ and decreasing it adiabatically (backward evolution),  we find a transition to the  $\text{CDW}_q$ phase at $V=V_q$. This phase remains stable down to small values of $V$, see Fig.~\ref{fig:small_U}-(d).  The change of momentum with $V$ is shown in Fig.~\ref{fig:small_U}-(f), which depicts the amplitude of the Fourier transform of the particle density. In the $\text{CDW}_q$  phase, the current decreases when decreasing $V$, see Fig.~\ref{fig:small_U}-(c).

The previous scenario contrasts with the forward evolution, where we start from $V=0$ and increase $V$. In this case, $\text{CDW}_\pi$  remains stable until $V$ bridges the equilibrium gap $V=\Delta$, see Fig.~\ref{fig:small_U}-(d). As expected, the current vanishes in the $\text{CDW}_\pi$  phase. Beyond $V=\Delta$, we observe an intermediate region bounded by $V_I$, for which the current increases with $V$ but is always smaller than in the $\text{CDW}_q$. However, in this regime the solutions of the mean-field self-consistent equations turn out to be meta-stable and delicately dependent on specific initial conditions. Fig.~\ref{fig:small_U}-(e) shows this meta-stable behavior for $\Delta<V<V_I$ in the Fourier transform of the charge density. This may signal that in this region, the mean-field solution cannot qualitatively capture the properties of the thermodynamic state. Therefore we will not be discussing this region any further in this work.
Beyond this intermediate region, for $V>V_I$, the $\text{CDW}_q$ phase becomes stable.

To further investigate the properties of the various phases, we show in Fig.~\ref{fig:small_U}-(g) and (h) the collective excitation gap, $\Delta_\text{coll}$, which corresponds to the smallest eigenvalue in absolute value of the inverse retarded RPA susceptibility $\left[ \bm{\chi}^R_\text{RPA}(\omega=0) \right]^{-1}$. The vanishing of this quantity signals the existence of collective eigenmodes of vanishing energies. 
$ \Delta_\text{coll}$  is finite in the disordered phase, {\itshape i.e.}, above $V_q$ and in the $\text{CDW}_\pi$ phase but  vanishes within the $\text{CDW}_q$ and in the intermediate phase.  
Within  $\text{CDW}_q$, the vanishing of $ \Delta_\text{coll} $ is consistent with the presence of a Goldstone mode ("phason") reflecting the degeneracy of phase excitations of an incommensurate CDW~\cite{Gruner88}.

The second-order phase transition line between $\text{CDW}_q$ and C was obtained by tracing the vanishing of $\Delta_\text{coll}$. We also verified that this is consistent with the appearance of oscillations in the density, with the correspondent wave vector, that are stable for increasing system sizes.

It is worth noting that the logarithmic divergence of $\chi_0^{R}[q_\text{max}(V)]$  with system size for $V<W$, implies that the voltage where the symmetric phase gets restored, $V_q$, is pushed to $V=W$  in the thermodynamic limit. However, for the system sizes available, we find $V_q \simeq 2$ with sizable finite size effects. Our finite size scaling analysis is rendered inconclusive by the slow dependence on $L$. 
Therefore, we expect that in the thermodynamic limit the mean-field analysis yields a stable $\text{CDW}_q$  within the full region marked as C in Fig.~\ref{fig:pd}-(b). 
On general grounds, one expects mean-field approaches to underestimate fluctuation effects that may suppress order, so that the beyond mean-field phase boundary of the putative $\text{CDW}_q$  is likely reduced below the bandwidth ($V=W$).

The inset of Fig.~\ref{fig:small_U}-(c) shows the current $J$ vs. $V$. In the insulating $\text{CDW}_\pi$  charge excitations are gapped and the current vanishes. Upon increasing $V$ further the  $\text{CDW}_\pi$ becomes unstable towards $q<\pi$ ordering. In the ensuing $\text{CDW}_q$ phase, $J$ becomes finite, with $\partial_V J>0$, and similarly for the $\text{C}$ phase.
Interestingly, within the $\text{MK-C}$ phase, which is obtained for voltages larger than the bandwidth, the differential conductance becomes negative, $\partial_V J<0$, as the current saturates to its infinite $V$ value. This negative differential conductance happens for a range of values of $U$.

\subsection*{Voltage driven CDW melting at strong coupling}

%%%%%%%%%%%%%%%%%%%%%%% Figure X %%%%%%%%%%%%%%%%%%%%%%%%%%
\begin{figure*}[t!]
\centering
\includegraphics[width= 1.0 \textwidth]{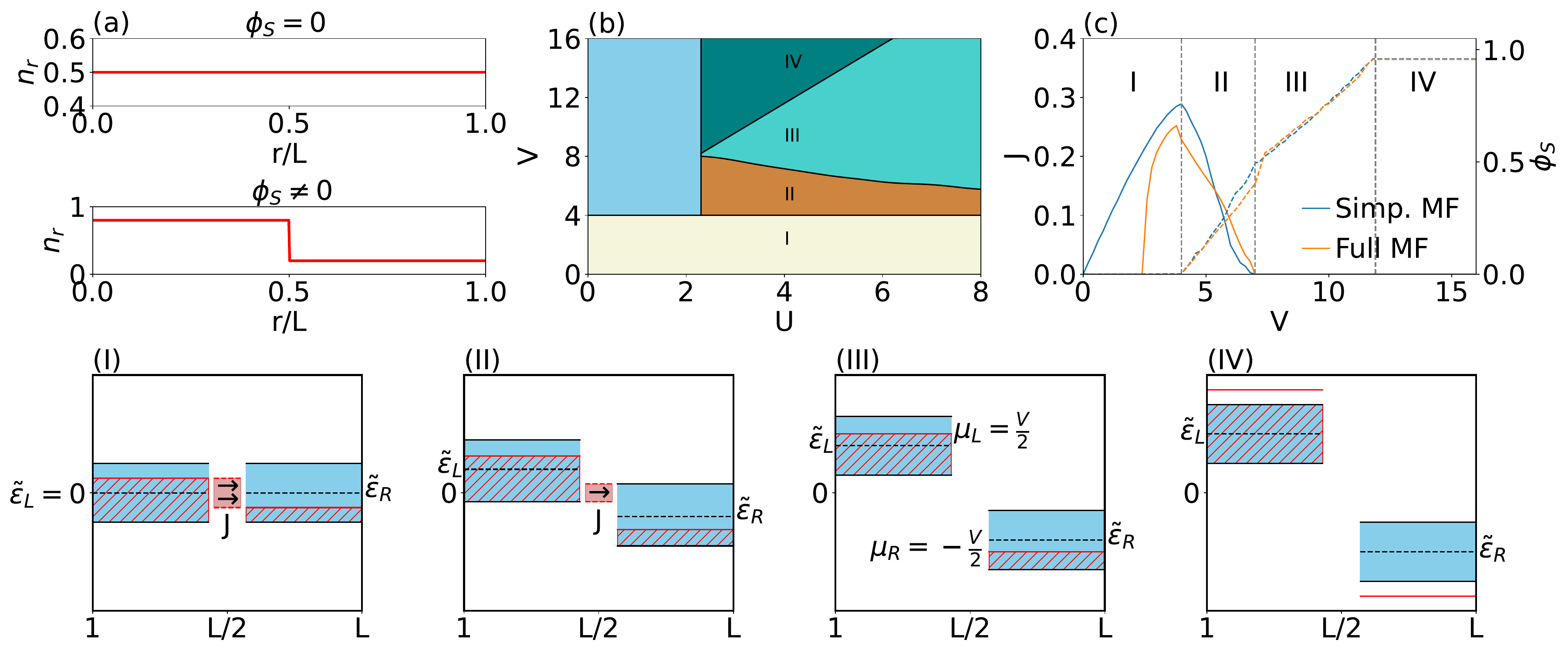} 
\caption{ Large coupling. (a) Schematic of the simplified mean-field approach comparing the density profiles of a vanishing and finite order parameter $\phi_S$. (b) Phase diagram of the simplified mean-field method showing the different regimes and transitions. (c) Current $J$ (continuous) and order parameter $\phi_S$ (dashed) as a function of $V$, comparing the simplified and full mean-field versions of these quantities. (I)-(IV) schematic of the bands for the different regimes. }
\label{fig:simp_model}
\end{figure*}
%%%%%%%%%%%%%%%%%%%%%%%%%%%%%%%%%%%%%%%%%%%%%%%%%%%%%%%%%%

Similarly to what happens in the weak-coupling regime, the $\text{CDW}_\pi $ at large $U$ is stable for an adiabatic increase of $V$ up to values of the charge gap $\Delta(U)$. 
For sufficiently large $U$,  upon further increase of $V$, the system transitions into another insulating regime, where the charge density assumes a step-like function with a domain wall pinned to the center of the chain separating the regions of low and high density. Within this region, the step order parameter, given by 
$\phi_{\text{S}} = \sum_{r=1}^{L} (2 n_r -1) \text{sgn}(L/2-r) / L$  (see Fig.~\ref{fig:simp_model}-(a)), is finite. We call this insulating phase with $0<\phi_{\text{S}}<1 $ the charge density step insulator (CDS-I). A further increase of  $V$ drives a continuous transition to a phase with a saturated $\phi_{\text{S}}$. This phase is adiabatically connected to the one previously established rigorously in the $V \rightarrow \infty$ limit~\cite{Prosen3,Prosen4}, where the dynamics is fully Markovian. We thus refer to it as the Markovian Insulating (MK-I) regime. 
 
Interestingly, the $\text{CDW}_\pi-\text{CDS-I}$ first order transition is foreshadowed by the appearance of boundary defects, marked in Fig.~\ref{fig:pd}-(b) by a purple dotted-dashed line (see representative density profile with and without defects in SM's Fig.3-(f) and (g) respectively~\cite{SupMat}-S4. A finite-size scaling analysis shows that  these defects are confined to the boundary and do not affect the $\text{CDW}_\pi$ bulk order parameter. Nevertheless, the effect is robust and corresponds to a divergence of the local susceptibility at RPA level (see SM~\cite{SupMat}-S5).

For a backward evolution, {\itshape i.e.}, decreasing $V$,  starting at the  $\text{CDS-I}$  phase, the system transitions continuously to a conducting phase with a charge density step ($\phi_{\text{S}}>0$ ). We denote this phase $\text{CDS-C}$. It can be stabilized down to $V=W$ regardless of the value of $U$. 

It is worth noting that $\text{CDS-C}$  is not realized in the forward evolution, when $V$ increases starting from the equilibrium  $\text{CDW}_\pi$. 
The metastable region where  $\text{CDW}_\pi$ and $\text{CDS-C}$ or $\text{CDS-I}$ coexist is represented in Fig.~\ref{fig:pd}-(b) with alternating stripes of the color of each of the phases. A similar convention is used for other metastable regions of the phase diagram. 

In order to analyze the physics of the strong coupling phases, we study a simplified mean-field model obtained by restricting the density configurations to follow a step-like configuration similar to the one obtained for large $V$, see SM~\cite{SupMat}-S7 for more details. 

In Fig.~\ref{fig:simp_model}-(b), we provide the phase diagram of this simplified model. The  applicability of this mean field model is confined to large $U$ where the density profiles are step-like. Therefore, we concentrate our analysis on the regions labelled I-to-IV in Fig.~\ref{fig:simp_model} which are in one-to-one correspondence with their counterparts in the full mean-field phase diagram. 

The current (continuous) and the order parameter (dashed) as a function of $V$ (in the backwards direction) are depicted in Fig.~\ref{fig:simp_model}-(c) for both the simplified and the full mean-field  methods. 

On either side of the charge density step, the system becomes translational invariant.  Thus, for  large system sizes, we can independently analyze the band structure of the left and right parts of the chain. A non-zero $\phi_{\text{S}}$ corresponds to a positive (negative) shift in the left (right) energy bands of magnitude $\tilde \varepsilon_{L/R} = \pm U \phi_{\text{S}}$.  
Fig.~\ref{fig:simp_model}-(I-IV) shows the renormalized energy bands of the translationally invariant system and the chemical potentials of the leads in different regions.

Phase I has $\phi_{\text{S}}=0$ and corresponds to a weekly interacting metal for which the differential conductance is positive $\partial_V J > 0$. Current is carried by electrons with energies  $\varepsilon \in [-V/2, V/2]$. This region is the counterpart of phase $\text{C}$,  which is not observed at large $U$  as it is unstable towards $\text{CDW}_\pi$ formation in this region. 

For Phase II, $\phi_{\text{S}}>0$, corresponding to a shift in the average band energy. Current is still flowing, carried by electrons with  $\varepsilon \in [\tilde \varepsilon_{L} - W/2, \tilde \varepsilon_{R} + W/2]$.
Increasing $V$ further increases $\phi_{\text{S}}>0$ and reduces the number of available carriers. As a consequence, one finds $\partial_V J < 0$. This region provides a simple physical picture of phase $\text{CDS-C}$. 

For Phase III, the renormalized right and left band energies no longer overlap  and the current vanishes as $\phi_{\text{S}}>0$ continues to increase with $V$. This region provides a simplified description of phase $\text{CDS-I}$. 

Finally, for Phase IV, the left (right) chemical potential reaches the top (bottom) of the corresponding band and the occupation saturates to unity (vanishes). This regions describes the features of the large $U$ MK-I phase.

\subsection*{Large Bias Regime}

%%%%%%%%%%%%%%%%%%%%%%% Figure 4 %%%%%%%%%%%%%%%%%%%%%%%%%%
\begin{figure}
\centering
\includegraphics[width= 0.5 \textwidth]{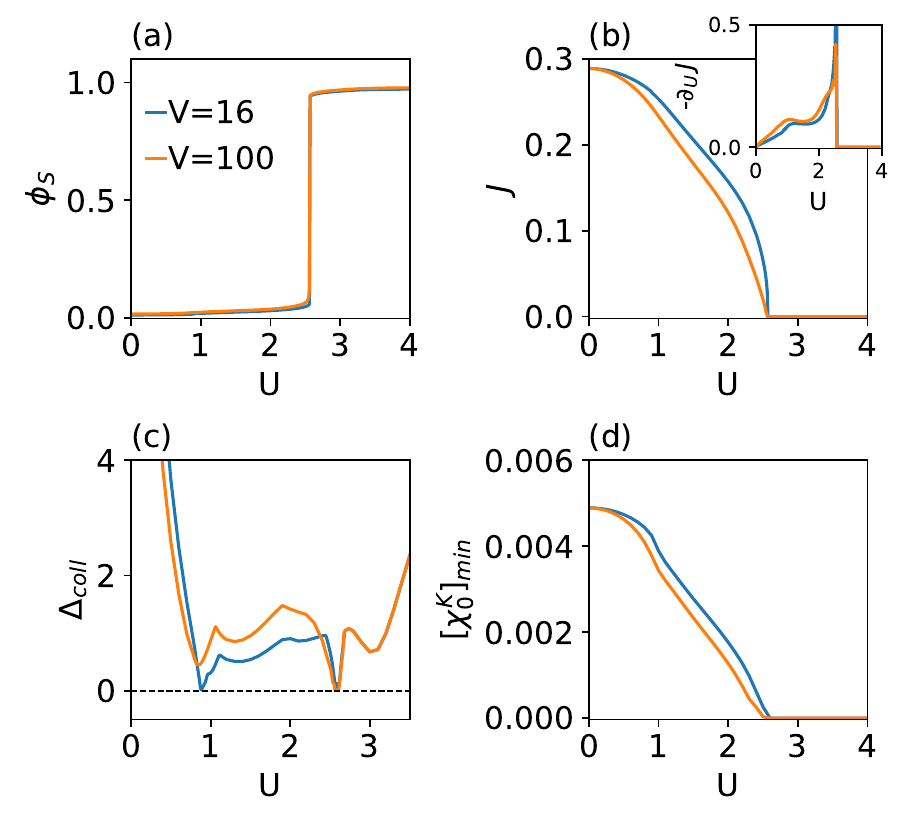} 
\caption{ Large bias. (a) Charge separated order parameter $\phi_{\text{S}}$ as a function of $U$, (b) current $J$ and differential conductivity $\partial_U \, J$ (inset), (c) collective excitation gap $\Delta_\text{coll}$ and (d) smallest eigenvalue amplitude of the Keldysh component of the bare susceptibility. These results are obtained for two different values of the bias $V=16$ and $V=100$. (a)-(c) correspond to system size of $L=100$ and (d) to $L=60$. } 
\label{fig:largeV}
\end{figure}
%%%%%%%%%%%%%%%%%%%%%%%%%%%%%%%%%%%%%%%%%%%%%%%%%%%%%%%%%%

This section discusses the large $V$ regime extending to the  Markovian limit for $V\to \infty$. 
As shown in Fig.~\ref{fig:pd}-(b),  for $V\gg U,W$  only the $\text{MK-C}$ and  $\text{MK-I}$  are realized.  However, for $V<V^* \simeq 29$ (see SM~\cite{SupMat}-S6), we find a line within the  $\text{MK-C}$ phase along which our RPA analysis predicts gapless collective modes, see Fig.~\ref{fig:largeV}-(c).  
As the two regions on either side of the line terminating in $V^*$ are adiabatically connected for large $V$, both regions form one phase. 
A precursor of this transition is seen in the current derivative when approaching  $V^*$ from above.  
Interestingly, even for $V\to \infty$, such a feature is observed in the current (see large $V$ in the inset of Fig.~\ref{fig:largeV}-(b) and $V\to \infty$ in the SM~\cite{SupMat}-S8). We note in passing that these mean-field results coincide with a marked change of unusually strong finite size corrections in the exact solution, see SM~\cite{SupMat}-S8.

Fig.~\ref{fig:largeV}-(a) shows the discontinuity of the $\phi_\text{S}$ order parameter at the $\text{MK-C}-\text{MK-I}$ transition concomitantly with the vanishing of the current, see Fig.~\ref{fig:largeV}-(b), and the vanishing of $\Delta_\text{coll}=0$, see Fig.~\ref{fig:largeV}-(c). The observation of a discontinuous order parameter accompanied by a divergent correlation length is reminiscent of the one encountered in the bias-driven order-disorder phase transition of the transverse-field Ising model~\cite{Tharnier1}. 

When $U$ takes on even larger values within the $\text{MK-I}$ phase, the collective gap reappears. However, the smallest eigenvalue amplitude of the Keldysh component of the bare susceptibility, $\chi_0^K(\omega=0)$, continues to vanish as $L\rightarrow \infty$ everywhere within the $\text{MK-I}$  phase, see Fig.~\ref{fig:largeV}-(d). 
The vanishing of the bare Keldysh component signals a decoupling of the corresponding mode from its environment, a feature sometimes referred to as dark mode~\cite{Marino_2016,Diehl_2008,Diehl_2010,Verstraete_2009}. These environment-decoupled modes are responsible for relaxation times that increase with $L$ and thus slow down the convergence to the steady-state as $L$ increases~\cite{Benenti1,Benenti2}.

\section*{Discussion}
The voltage-driven breakdown of order is a topic of multifaceted implications, from theoretical understanding to practical applications. 
Our analysis demonstrates a perhaps unexpectedly rich interplay between bias and interactions, which occurs already on the level of archetypical models.

For the specific model  considered here, a charge-density wave system, our mean-field analysis uncovers a plethora of non-trivial phases including some that are adiabatically connected to well established limits in voltage ($V=0$, $V\to\infty$) and interaction strength ($U=0$, $U\to \infty$). 

Breaking the insulating phase requires a voltage of the order of the equilibrium quasi-particle gap. We find that this insulator-breakdown phase transition occurs is accompanied by a discontinuity of the order parameter. Whether this is a non-equilibrium mixed-order transition, that is accompanied by a divergent correlation length, similar to the one previously found in refs.~\cite{Tharnier1,Tharnier2}, is an interesting question that requires further analysis beyond the current mean-field analysis.    

Once the order is destroyed, the hysteretical nature of the system allows us to investigate the phases beneath the gap by adiabatically reducing the voltage. 

At weak coupling and finite voltage, we found that these hidden phases  include a current-carrying CDW-ordered phase with a voltage-dependent wave-vector that transitions to a disordered state upon further increasing the voltage.
The precise nature of this transition and its universal features are an important open issue that we hope to address in future studies. 

In the strong-coupling limit, we present a simplified analysis, which reproduces the mean-field phase diagram and provides a clear physical picture. 
In particular, this model explains the negative conductance observed across a conducting charge-separated phase and the subsequent transitions to a succession of insulating phases upon increasing the voltage. 

Our RPA analysis of the strong-coupling insulating phase shows the existence of dark modes in the  large voltage phase which addiabatically connects to the Markovian regime.
These modes consist of gaped excitations of the domain wall separating the different charge density regions whose dynamics decouple from the environment.

Faithfully capturing the weak-coupling regime in low-dimensional systems is challenging for any mean-field method, already in equilibrium. Thus, while the results reported here in this limit should be interpreted with care, the finite-bias results give a valuable qualitative picture that can guide further research. 
Specifically, the stability of the $\text{CDW}_q$ phase, 
which we find only away from equilibrium, remains uncertain once quantum fluctuations are taken into account. 
In one-dimension, under equilibrium conditions, such a phase is prohibited by the Mermin-Wagner theorem.
However, no extension of Mermin-Wagner's theorem to out-of-equilibrium phase stability is known as of now. Nevertheless, for some classical systems, non-equilibrium violations of the Mermin-Wagner theorem were observed to stabilize order~\cite{Toner1995,Galliano2023}. 

If the $\text{CDW}_q$ phase does not survive the inclusion of quantum fluctuations in one dimension, it would be interesting to understand what replaces it.
Irrespectively of these considerations, it is conceivable that 
$\text{CDW}_q$ could be stabilized in quasi-one-dimensional materials with weak transversal couplings, as has been demonstrated to occur in equilibrium~\cite{Duck2022}. 
Indeed, the mean-field approach is expected to work better in higher space dimensions. Thus, it will be interesting to see if this method can faithfully capture the large-bias regimes in two dimensions, where no exact solution is yet known.

\section*{Acknowledgements}

M.O. acknowledges support by FCT through Grant No. SFRH/BD/137446/2018 and BL194/2023\_IST-ID.
M.O. and P.R. acknowledge further support from FCT-Portugal through Grant No. UID/CTM/04540/2020, and from DQUANT QuantERA II Programme  through FCT-Portugal Grant Agreement No. 101017733. S.K. acknowledges support by the National Science and Technology Council of Taiwan through Grant No.\ 112-2112-M-A49-MY4, the Yushan Fellowship Program of the Ministry of Education of Taiwan and the Center for Theoretical and Computational Physics of NYCU, Taiwan.
Computations were performed in the Bob MACC cluster through the Advanced Computing Project CPCA/A1/470166/2021.

\bibliographystyle{apsrev4-1}
\bibliography{References}

\newpage 

\begin{widetext}
\begin{center}
\textbf{\large{}\textemdash{} Supplemental Material \textemdash{}}
\par\end{center}{\large \par}
\begin{center}
\textbf{\large{}Voltage-Driven Breakdown of Electronic Order}
\par\end{center}{\large \par}
\begin{center}
\textbf{Miguel M. Oliveira$^{(a)}$, Pedro Ribeiro$^{(a,b)}$, and Stefan Kirchner$^{(c)}$}\\
$^{(a)}$CeFEMA, Instituto Superior T\'ecnico, Universidade de Lisboa Av. Rovisco
Pais, 1049-001 Lisboa, Portugal\\
$^{(b)}$Beijing Computational Science Research Center, Beijing 100193, China\\
$^{(c)}$Department of Electrophysics, National Yang Ming Chiao Tung University, Hsinchu 30010, Taiwan
\end{center}
\begin{description}
\item [{Summary}] Below we provide additional technical details and further numerical results supplementing the conclusions from the main text.
\end{description}
\end{widetext}

\beginsupplement

\section{Keldysh Action \label{Keldysh Action}}

In this section a self-contained derivation of Eqs.~(\textcolor{blue}{4,6,8-11})
of the main text is provided.

We start with the non-interacting case, where the Keldysh action is given by 
\begin{equation}
Z = \int \mathcal{D}[\bar{\bm{\Psi}}, \bm{\Psi}] \, e^{i S_0[\bar{\bm{\Psi}}, \bm{\Psi}] }, 
%= \int \mathcal{D}[\bar{\bm{\Psi}}, \bm{\Psi}] \, e^{i \, \int_{\text{C}} dz dz' \bar{\bm{\Psi}}(z) \, \bm{G}^{-1}_0(z,z') \, \bm{\Psi}}(z')  \sk{.}
\label{Z0}
\end{equation}
here 
\begin{equation}
 S_0 = \int_{\text{C}} dz dz' \bar{\bm{\Psi}}(z) \, \bm{G}^{-1}_0(z,z') \, \bm{\Psi}(z'),
\end{equation}
is the Keldysh action integrated over the contour of Fig.~\ref{contour} and $\bm{\Psi}= \begin{pmatrix} \bm{\Psi}_C & \bm{\Psi}_R & \bm{\Psi}_L \end{pmatrix} ^T$ represents the spinless fermion fields,
where the  components $\bm{\Psi}_C$, and $\bm{\Psi}_R$, and $\bm{\Psi}_L$ represent the chain, and right and left reservoir respectively.
The fields $\bm{\Psi}_{C,r}(z)$ have a label $r$ corresponding to the lattice site and a parameter $z$ indicating the position on the Keldysh contour $\mathcal{C}$. For details of the Keldysh technique  see {\itshape e.g.} Ref.~\cite{kamenev2011}. $\bm{G}^{-1}_0$ is the  inverse non-interacting Green function and is given by 
\begin{equation}
\bm{G}^{-1}_0 = 
\begin{pmatrix}
\bm{G}^{-1}_{C,0} & - \bm{T}_R & -\bm{T}_L \\
-\bm{T}^\dagger_R & \bm{G}^{-1}_{R,0} & 0 \\
-\bm{T}^\dagger_L & 0 & \bm{G}^{-1}_{L,0}
\end{pmatrix} \quad,
\end{equation}
where $\bm{G}^{-1}_{C,0}$ is the inverse Green function of the chain and $\bm{G}^{-1}_{l,0}$ is that of the l'th reservoir; $\bm{T}_l$ encodes the coupling between the chain and the reservoirs. These components have the following definitions
\begin{align}
&\bm{G}^{-1}_{C,0} (z,z') = \delta(z-z') \left( i \partial_{z'} - \bm{H}_C \right) \\
&\bm{G}^{-1}_{l,0} (z,z') = \delta(z-z') \left( i \partial_{z'} - \bm{H}_l \right) \\
& \bm{T}_l(z,z') = \delta(z-z') \, \theta(z-t_0) \, \bm{J}_l \quad ,
\end{align}
where $\bm{H}_C$ is the single-particle Hamiltonian of the chain and $\bm{H}_l$ describes the dynamics of the l'th reservoir. Here, $\bm{J}_L$ and $\bm{J}_R$ are matrices that determine the chain-reservoir coupling in accordance with Eq.~(\textcolor{blue}{3})
of the main text and are given by
\begin{equation}
\bm{J}_L = 
\begin{pmatrix}
\frac{J_L}{\sqrt{N}} & \cdots & \frac{J_L}{\sqrt{N}} \\
0 & \cdots & 0 \\
\vdots & & \vdots \\
0 & \cdots & 0
\end{pmatrix} \quad \text{and \quad}
\bm{J}_R = 
\begin{pmatrix}
0 & & 0 \\
0 & \cdots & 0 \\
\vdots & & \vdots \\
\frac{J_R}{\sqrt{N}} & \cdots & \frac{J_R}{\sqrt{N}}
\end{pmatrix}  .
\end{equation}

By integrating out the reservoirs from Eq.~\eqref{Z0}, one obtains an effective action that depends only on the chain fields and is given by $S_0\left[ \bar{\bm{\Psi}}_C , \bm{\Psi}_C \right] = \bar{\bm{\Psi}}_C \, \bm{G}^{-1}_C \, \bm{\Psi}_C$ with $\bm{G}^{-1}_C = \bm{G}^{-1}_{C,0} - \bm{\Sigma}_C$.
The self-energy of the chain is $\bm{\Sigma}_C= \sum_l \bm{T}_l \, \bm{G}_{l,0} \, \bm{T}^\dagger_l$.
For clarity, we drop the chain label from the fields in what follows below, {\itshape i.e.}, $\bm{\Psi}_C \equiv \bm{\Psi}$ and likewise for the Green function.

\subsection*{Non-equilibrium propagators}

%%%%%%%%%%%%%%%%%%%%%%% Contour %%%%%%%%%%%%%%%%%%%%%%%%%%
\begin{figure}[t!]
\centering
\includegraphics[width= 0.5 \textwidth]{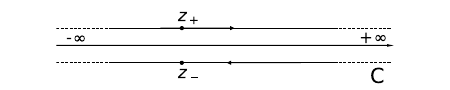} 
\caption{ Image of the Keldysh contour, showcasing the forward and backward branch.}
\label{contour}
\end{figure}
%%%%%%%%%%%%%%%%%%%%%%%%%%%%%%%%%%%%%%%%%%%%%%%%%%%%%%%%%%

The fermion fields are defined on the Keldysh contour, with $\Psi^+(t)$ and $\Psi^-(t)$ labeling fields on the forward and backward branch respectively, see Fig.~\ref{contour}. The variable $z$ is used to parameterize a generic point on the Keldysh contour according with $\Psi^+(t) = \Psi^+(z=t +i0)$ and $\Psi^-(t) = \Psi^+(z=t -i0)$. The Green function has a corresponding structure given by
\begin{equation}
\bm{G}(t,t') = 
\begin{pmatrix}
\bm{G}^\mathbb{T}(t,t') & \bm{G}^<(t,t')  \\
\bm{G}^>(t,t') & \bm{G}^{\tilde{\mathbb{T}}}(t,t')  \\
\end{pmatrix} \quad,
\end{equation}
where the different components are defined in terms of averages over Grassmann fields as
\begin{align}
G_{rr'}^<(t,t') &= -i \, \langle \Psi_r^+(t) \, \bar{\Psi}_{r'}^-(t') \rangle \nonumber \\
G_{rr'}^>(t,t') &= -i \, \langle \Psi_r^-(t) \, \bar{\Psi}_{r'}^+(t') \rangle \nonumber \\
G_{rr'}^\mathbb{T}(t,t') &= -i  \, \langle \Psi_r^+(t) \, \bar{\Psi}_{r'}^+(t') \rangle \nonumber \\
& = \theta(t-t') \, G_{rr'}^>(t,t') + \theta(t'-t) \, G_{rr'}^<(t,t') \nonumber \\
G_{rr'}^{\tilde{\mathbb{T}}}(t,t') &= -i \, \langle  \Psi_r^-(t) \, \bar{\Psi}_{r'}^-(t') \rangle \nonumber \\
& = \theta(t'-t) \, G_{rr'}^>(t,t') + \theta(t-t') \, G_{rr'}^<(t,t') \quad,
\end{align}
where $\theta(t-t')$ is the Heaviside step function and
each of the components corresponds to a different permutation of the fields on the two branches.

Due to the causal structure, the four 
Green function components are not linearly independent, as can be seen from $\bm{G}^\mathbb{T}(t,t') + \bm{G}^{\tilde{\mathbb{T}}}(t,t') - \bm{G}^<(t,t') - \bm{G}^>(t,t') = 0$, (For the subtleties associated with  $t=t'$, see Ref.~\cite{kamenev2011}). It is thus convenient to perform the following {\itshape Keldysh rotation}
\begin{align}
\bm{\Psi}_1(t) &= \frac{ \bm{\Psi}^+(t) + \bm{\Psi}^-(t)}{\sqrt{2}}, \quad \quad \bm{\Psi}_2(t) = \frac{ \bm{\Psi}^+(t) - \bm{\Psi}^-(t)}{\sqrt{2}}, \nonumber \\
{\bm{\bar\Psi}}_1(t) &= \frac{ {\bm{\bar\Psi}}^+(t) - {\bm{\bar\Psi}}^-(t)}{\sqrt{2}}, \quad \quad {\bm{\bar\Psi}}_2(t) = \frac{ {\bm{\bar\Psi}}^+(t) + {\bm{\bar\Psi}}^-(t)}{\sqrt{2}} .
\label{rotation}
\end{align}
This transformation yields a more compact form for the Green function
\begin{equation}
\bm{G}(t,t') = 
\begin{pmatrix}
\bm{G}^R(t,t') & \bm{G}^K(t,t')  \\
0 & \bm{G}^A(t,t')  \\
\end{pmatrix} \quad,
\end{equation}
where $\bm{G}^R$, $\bm{G}^A$ and $\bm{G}^K$ are called the retarded, advanced and Keldysh Green function respectively. They are defined as
\begin{align}
\bm{G}^R(t,t') &= \theta(t-t') \left[ \bm{G}^>(t,t') - \bm{G}^<(t,t') \right] \nonumber \\
\bm{G}^A(t,t') &= \theta(t'-t) \left[ \bm{G}^<(t,t') - \bm{G}^>(t,t') \right] \nonumber \\
\bm{G}^K(t,t') &= \bm{G}^<(t,t') + \bm{G}^>(t,t') \quad ,
\end{align}
and obey the following symmetries $\left[ \bm{G}^{<(>)} \right]^\dagger = -\bm{G}^{<(>)}$, $\left[ \bm{G}^R \right]^\dagger = \bm{G}^A$ and $\left[ \bm{G}^K \right]^\dagger = -\bm{G}^K$. Here the Hermitian transpose is understood as a complex conjugation followed by a transposition in the time and lattice labels.

The inverse Green function and self-energy have a matching causal structure given by
\begin{equation}
\bm{G}^{-1} = 
\begin{pmatrix}
\left[\bm{G}_0^{R}\right]^{-1} - \bm{\Sigma}^R & \left[\bm{G}_0^{-1}\right]^K - \bm{\Sigma}^K \\
0 & \left[\bm{G}_0^{A}\right]^{-1} - \bm{\Sigma}^A \\
\end{pmatrix} \quad.
\label{Inverse_Green}
\end{equation}
Note that $\left[\bm{G}_0^{-1}\right]^K$ is the Keldysh component of the inverse Green function and not the inverse of the Keldysh component; nonetheless it amounts to just a regularization of the non-interacting case and can be neglected here due to the presence of a finite Keldysh self-energy. As for the retarded and advanced components we have that $\left[\bm{G}_0^{-1}\right]^{R(A)} = \left[\bm{G}_0^{R(A)}\right]^{-1}$, which is a consequence of the necessity for $\bm{G}_0^{-1}$ to be the inverse of $\bm{G}_0$.

Using the Dyson equation $\bm{G}^{-1} \circ \bm{G} = \bm{1}$, we can obtain each of the components of the Green function. Here the symbol $\circ$ represents a convolution in time, space and the causal structure of the matrices. This yields the following equation for each of the components
\begin{align}
&\left( \bm{G}_0^{R-1} - \bm{\Sigma}^R \right) \circ \bm{G}^R = \bm{1} \nonumber \\
&\left( \bm{G}_0^{A-1} - \bm{\Sigma}^A \right) \circ \bm{G}^A = \bm{1} \nonumber \\
&\left( \bm{G}_0^{R-1} - \bm{\Sigma}^R \right) \circ \bm{G}^K - \bm{\Sigma}^K \circ \bm{G}^A = 0 \quad.
\end{align}
As was said earlier, we are only interested in studying steady-states; for which time-translation invariance is assumed. Carrying out a Fourier transformation yields eqs.~(\textcolor{blue}{6})
from the main text.

\subsection*{Wide-band Limit}

Currently we have in Eq.~(\textcolor{blue}{6})
of the main text an expression for each component of the Green function written in terms of the self-energies, which now have to be determined. In order to do this we use the wide-band limit, which consists in assuming that the energy bandwidth of the reservoirs is much wider than that of the system. From the perspective of the chain, this assumption effectively turns the density of states of the reservoirs into a constant $\rho_l(\epsilon) = \frac{1}{N}\sum_k \delta(\epsilon-\epsilon_{l,k}) \approx \rho_l$. We also assume that the coupling between the chain and the reservoirs was turned on in the infinite past $t_0 \to -\infty$ and that reservoirs start and remain in thermal equilibrium the whole time.

 Starting from the previous result $\bm{\Sigma} = \sum_l \bm{T}_l \, \bm{G}_{l,0} \, \bm{T}^\dagger_l$, we can obtain the following for the retarded and advanced components in frequency space
\begin{align}
& \left[ \bm{\Sigma}^{R/A}(\omega) \right]_{r,r'} = \sum_l \sum_{k,k'} \left[ \bm{J}_l \right]_{r,k} \, \left[ \bm{G}^{R/A}_{l,0}(\omega) \right]_{k,k'} \, \left[ \bm{J}^\dagger_l \right]_{k',r'} \nonumber \\
&= \mp i\pi \sum_l J_l^2 \, \delta_{r,l} \, \delta_{l,r'} \, \rho_l = \mp i \sum_l \frac{\left[ \Gamma_l \right]_{r,r'} }{2} \quad ,
\label{self-retarded-freq}
\end{align}  
where we have used the equilibrium Green functions for the reservoirs $G^{R/A}_{l,0} =(\omega - \epsilon_{l,k} \pm i 0 )^{-1}$. We also took the principal value integral to vanish. In matrix notation this result is written as $\bm{\Sigma}^{R/A}(\omega) = \mp i \sum_l \bm{\Gamma}_l / 2$, where $\bm{\Gamma}_l = 2\pi \rho_l \bm{J}_l \bm{J}_l^\dagger$.

Likewise, for the Keldysh component of the self-energy we have 
\begin{align}
& \left[ \bm{\Sigma}^K(\omega) \right]_{r,r'} = \sum_l \sum_{k,k'} \left[ \bm{J}_l \right]_{r,k} \, \left[ \bm{G}^K_{l,0} (\omega) \right]_{k,k'} \, \left[ \bm{J}^\dagger_l \right]_{k',r'} \nonumber \\
&= -i \sum_l \tanh \left[ \frac{\beta_l (\omega-\mu_l)}{2} \right] \left[\bm{\Gamma}_l \right]_{r,r'} \quad ,
\label{self-Keldysh-freq}
\end{align}
where once again we used the equilibrium Keldysh Green function $\bm{G}^K_{l,0} (\omega) = -2\pi i \left[ 1 - 2 f_l(\omega) \right] \delta(\omega-\epsilon_{l,k})$. Here $f_l(\omega) =  \frac{1}{e^{\beta_l(\omega-\mu_l)}+1}$ is the Fermi-Dirac distribution; and $\beta_l$ and $\mu_l$ are respectively the inverse temperature and chemical potential of the $l$'th reservoir.

A Fourier transformation of eqs.~\eqref{self-retarded-freq} and \eqref{self-Keldysh-freq} yields the result in Eq.~(\textcolor{blue}{4})
of the main text.

\subsection*{Non-interacting Susceptibility}

In this section, we indicate the derivation of the non-interacting susceptibilities in the Keldysh formalism. The presentation is based on Ref.~\cite{kamenev2011}. In what follows we find it convenient to use the variable $x$ to refer to the combined coordinate of lattice site $r$ and time variable $t$, \textit{i.e.}, $x=(r,t)$ and $\int dx \equiv \int dt \sum_r$.

The density-density susceptibility $\bm{\chi}_0$ can be obtained by considering an external field $J_r(t)$ that couples to the fermion density and then performing derivatives of the generating function with respect to that field, which in the end is set to zero. Previously we introduced the non-interacting action $S_0$ and now we consider the action of the coupling with the external field given by
\begin{equation}
S_J = -\int_C dz \sum_r J_r(z) \, \bar{\psi}_r(z) \, \psi_r(z) \quad .
\end{equation}
It will also be useful to introduce the Keldysh rotation of the external field, such that $J^+ = J^{cl} + J^q$ and $J^- = J^{cl} - J^q$. With this transformation and the corresponding one for the electron fields, shown in Eq.~\eqref{rotation}, we can rewrite the external field action 
\begin{align}
S_J &= \nonumber \\
& -\int_{-\infty}^\infty dt \sum_r \left[ J^+_r(t) \, \bar{\psi}^+_r(t) \, \psi^+_r(t) - J^-_r(t) \, \bar{\psi}^-_r(t) \, \psi^-_r(t) \right] \nonumber \\
&= -\int dx \sum_{a,b=1}^2 \bar{\psi}_a(x) \left[ J^{cl}(x) \, \gamma^{cl}_{ab}  + J^q(x) \, \gamma^q_{ab}  \right] \psi_b(x) \quad ,
\label{SJ}
\end{align}
where we introduced the matrices 
\begin{equation}
\bm{\gamma}^{cl} =
\begin{pmatrix}
1 & 0 \\
0 & 1
\end{pmatrix}
\quad \text{and} \quad
\bm{\gamma}^{q} =
\begin{pmatrix}
0 & 1 \\
1 & 0
\end{pmatrix} \quad .
\label{gamma_matrices}
\end{equation}
Putting all of this together we have for the total action
\begin{align}
& S= S_0 + S_J \nonumber \\ 
& = \int dx \, dx'
\begin{bmatrix}
\bar{\psi}_1(x) & \bar{\psi}_2(x)
\end{bmatrix} 
\left[ \bm{G}^{-1} - J^\alpha \, \bm{\gamma}^\alpha \right] (x,x')
\begin{bmatrix}
\psi_1(x') \\ 
\psi_2(x')
\end{bmatrix} ,
\label{S0}
\end{align}
where we consider a Einstein sum over dummy indices such that $J^\alpha \, \bm{\gamma}^\alpha = J^{cl} \, \bm{\gamma}^{cl} + J^q \, \bm{\gamma}^q$. Note also that when we wrote the full action in Eq.~\eqref{S0}, out of convenience we took $J^\alpha$ to be a diagonal matrix in the space-time coordinate, i.e. $J^\alpha(x,x') = J^\alpha(x) \, \delta(x-x')$. 

We can now perform the Gaussian integration of the fermion fields leading to
\begin{align}
Z[J] &= \frac{\det \left[ i \, \bm{G}^{-1} - i \, J^\alpha \, \bm{\gamma}^\alpha \right] }{\det \left[ i \, \bm{G}^{-1} \right]} \nonumber \\ 
&= \det \left[ \bm{1} - \bm{G} \, J^\alpha \, \bm{\gamma}^\alpha \right] = e^{\Tr \, \log \left[ \bm{1} - \bm{G} \, J^\alpha \, \bm{\gamma}^\alpha \right]} \quad .
\end{align}
The non-interacting susceptibility is now obtained by differentiating the generating function $Z[V]$ with respect to the external field, see Ref.~\cite{kamenev2011} for more details. We thus obtain
\begin{align}
\chi^{\alpha \beta}_0 (x,x') &= -\frac{i}{2} \left. \frac{\delta^2 \log Z[J]}{ \delta J^\beta(x') \, \delta J^\alpha(x) } \right|_{J=0} \nonumber \\
&= \frac{i}{2} \Tr \left[ \bm{\gamma}^\alpha \, \bm{G}(x,x') \, \bm{\gamma}^\beta \, \bm{G}(x',x) \right] \quad ,
\label{Pol0_def}
\end{align}
where here the trace is only over the causal structure of the Green's functions. The structure of the susceptibility matrix is 
\begin{equation}
\bm{\chi}_0 (x,x')=
\begin{pmatrix}
0 & \chi^A(x,x') \\
\chi^R(x,x') & \chi^K(x,x')
\end{pmatrix} \quad ,
\label{Pol0_struct}
\end{equation}
which is the same causal structure as that of a bosonic self-energy. From Eq.~\eqref{Pol0_def} we can now obtain the different components of the susceptibility and since we are focused in the steady-state, we take the response functions to depend only on time differences, which allows us to go into frequency space. The same cannot be done for space since we are dealing with open-boundary conditions with reservoirs attached to the edges, which explicitly break translational invariance. Performing a Fourier transformation yields eqs.~(\textcolor{blue}{10}) and (\textcolor{blue}{11}) of the main text.

\subsection*{Mean-Field Self-Consistent Equations}

In this section we consider the full interacting problem and obtain the self-consistent mean-field equations. Below we derive the RPA expression for the susceptibility for the model considered in the main text.  

We start by separating the interacting Hamiltonian into the quartic and quadratic parts
\begin{align}
H_\text{int} &= U \sum_{r=1}^{L-1} \left( c^\dagger_r \, c_r - \frac{1}{2} \right) \left( c^\dagger_{r+1} \, c_{r+1} - \frac{1}{2} \right) \nonumber \\
&= \frac{1}{2} \sum_{r,r'} U_{r r'} \, \rho_r \, \rho_{r'} - \frac{U}{2} \sum_{i=1}^{L-1} \left( c^\dagger_r \, c_r + c^\dagger_{r+1} \, c_{r+1} \right) + \frac{U}{4}  ,
\end{align} 
where we introduced a density operator, defined as $\rho_r = c^\dagger_r \, c_r$, and the matrix $U_{r,r'}$, which in this case is symmetric and off-diagonal, accounting for the nearest-neighbor interactions. The quadratic terms are now included in the non-interacting Hamiltonian $H_0$ and from now on we deal with the interacting part alone. The interacting part of the action then takes the form
\begin{align}
& S_{int} = -\frac{1}{2} \int_C dz \sum_{r,r'} \rho_r(z) \, U_{r r'} \, \rho_{r'}(z) \nonumber \\
&= -\frac{1}{2} \int_{-\infty}^\infty dt \sum_{r,r'} \left[ \rho_r^+(t) \, U_{r r'} \, \rho^+_{r'}(t) - \rho_r^-(t) \, U_{r r'} \, \rho^-_{r'}(t) \right] .
\label{interacting_action}
\end{align}
We now consider a Hubbard–Stratonovich (HB) transformation to decouple the fermion interaction, introducing a bosonic real field. The transformation is based on the following identity
\begin{align}
& e^{-\frac{i}{2} \int_C dt \, \sum_{r,r'} \rho_r(t) \, U_{r r'} \, \rho_{r'}(t)} = \nonumber \\
& \int \mathcal{D}\varphi \, e^{i \int_C dt \left[ \frac{1}{2} \sum_{r,r'} \varphi_r(t) \, U^{-1}_{r r'} \, \varphi_{r'}(t) - \sum_r \varphi_r(t) \, \rho_r(t) \right] } \quad ,
\label{interacting}
\end{align}
where $\varphi_r(t)$ is a real bosonic field and the matrix $U^{-1}_{r r'}$ is the inverse matrix of $U_{r r'}$, not the inverse of the coefficients. The equality results from Gaussian integration, see for example Ref.~\cite{kamenev2011}, and $\mathcal{D}\varphi$ includes the normalization factor that makes the equality true.

The HB fields can now be Keldysh rotated according to 
\begin{align}
& \varphi^+ \, U^{-1} \, \varphi^+ - \varphi^- \, U^{-1} \, \varphi^- \nonumber \\
&= 2 \, \varphi^{cl} \, U^{-1} \, \varphi^q + 2 \, \varphi^q \, U^{-1} \, \varphi^{cl} = 2 \, \varphi^\alpha \, U^{-1} \, \sigma_x^{\alpha \beta} \, \varphi^\beta \quad ,
\end{align} 
where we dropped the $r$ and $t$ labels for simplicity. $\sigma_x$ is the Pauli matrix and in the last equality we assumed Einstein's summation notation over repeated indices. The second term in the action shown in Eq.~\eqref{interacting} has the same form as that for the external field $J$, so we deal with it in the same way. We can then write the full action of the model as
\begin{align}
S &= \int dx \, dx' \bar{\bm{\psi}}(x) \left[ \bm{G}^{-1} - J^\alpha \, \bm{\gamma}^\alpha - \varphi^\alpha \, \bm{\gamma}^\alpha \right] (x,x') \, \bm{\psi}(x') \nonumber \\
& \quad + \int dx \, dx' \, \bm{\varphi}^T(x) \, U^{-1}(x,x') \, \bm{\sigma}_x \, \bm{\varphi}(x') \quad ,
\end{align}
where, as before, we are taking $\varphi^\alpha(x,x') = \varphi^\alpha(x) \, \delta(x-x') = \varphi_r^\alpha(t) \, \delta_{r,r'} \, \delta(t-t')$ and $U^{-1}(x,x') = U^{-1}_{r,r'} \, \delta(t-t')$. Also we have $\bm{\varphi}= (\varphi^{cl} \, \, \varphi^q)^T$ and $\bm{\psi}= (\psi_1 \, \, \psi_2)^T$.

The action is now quadratic in the fermion fields, which can be integrated out to obtain
\begin{align}
S[J,\varphi] &= \int \int dx \, dx' \, \varphi^\alpha(x) \, U^{-1}(x,x') \, \sigma^{\alpha \beta}_x \, \varphi^\beta(x') \nonumber \\ 
& \quad - i \, \Tr \, \log \left \{ \bm{1} - \bm{G} \left[ \bm{J}^\alpha + \bm{\varphi}^\alpha \right] \bm{\gamma}^\alpha \right \} \quad ,
\label{S_J_phi}
\end{align}
which depends only on the external field $J$ and the HB field $\varphi$. The mean-field solution is obtained by minimizing the action with respect to the HB field such that $\frac{\delta S}{\delta \varphi^\alpha(x)}=0$ and sending the external field to zero. Doing so yields the solution
\begin{equation}
 \varphi^\alpha(x) = -\frac{i}{2} \, \sigma^{\alpha \beta}_x \int dx' \, U(x,x') \Tr \left[ \tilde{\bm{G}}(x',x') \, \bm{\gamma}^\beta \right] \quad ,
\end{equation}
where we introduced the mean-field Green function given by $\tilde{\bm{G}} = \left( \bm{G}^{-1} - \bm{\varphi}^\gamma \, \bm{\gamma}^\gamma \right)^{-1}$. Note also that the trace is now only over the causal structure of the Green's function. Specifying the result for each of the components we get 
\begin{align}
& \varphi^q(x) = 0 \nonumber \\
& \varphi^{cl}(x) = -\frac{i}{2} \int dx' \, U(x,x') \, \tilde{G}^K(x',x') \nonumber \\
& \quad \quad \quad = -\frac{i}{2} \sum_{r'} U_{r,r'} \left[ \tilde{G}^K \right]_{r',r'}(t,t) \quad ,
\label{MF_eq}
\end{align}
which can be further simplified in what is shown in Eq.~(\textcolor{blue}{5}) of the main text. This constitutes the mean-field self-consistent equation. We are only interested in the long-time regime for which the system has converged to a steady-state, so we can drop the explicit time dependence from the field. In this limit the system also becomes translation invariant in time, so for completeness sake we write the mean-field Green's functions in frequency space
\begin{align}
 \tilde{\bm{G}}^R (\omega) &= \left[ \bm{G}^{R-1}(\omega) - \bm{\varphi}^{cl} \right]^{-1} \nonumber \\
 \tilde{\bm{G}}^A (\omega) &= \left[ \bm{G}^{A-1}(\omega) - \bm{\varphi}^{cl} \right]^{-1} \nonumber \\
 \tilde{\bm{G}}^K (\omega) &= \tilde{\bm{G}}^R(\omega) \, \bm{\Sigma}^K(\omega) \, \tilde{\bm{G}}^A(\omega)  \quad .
\end{align}
The method then works as follows: starting from initial field configuration $ \varphi^{cl}_0$ we compute the Green functions and then we use Eq.~\eqref{MF_eq} to update the fields. This procedure is iterated on until convergence is achieved.

\subsection*{RPA Susceptibility}

Here we provide a sketch of the derivation of the RPA susceptibility, which was used in the main text to study the collective excitations present given a certain mean-field solution $\varphi_\text{MF}$.

We start by rewriting the action in Eq.~\eqref{S_J_phi} under the change of variables $\varphi \to \varphi + \varphi_\text{MF}$, such that now $\varphi$ represents fluctuations around the mean-field solution $\varphi_\text{MF}$,
\begin{align}
& S[J,\varphi] \nonumber \\ 
&= \int dt \sum_{r,r'} \left[ \bm{\varphi}_r^T(t) + \bm{\varphi}_{\text{MF}, r}^T \right] U^{-1}_{rr'} \, \bm{\sigma}_x \left[ \bm{\varphi}_{r'}^T(t') + \bm{\varphi}_{\text{MF}, r'}^T \right] \nonumber \\
& \quad  - i \, \Tr \, \log \left \{ \bm{1} - \tilde{\bm{G}} \left[ \bm{J}^\alpha + \bm{\varphi}^\alpha \right] \bm{\gamma}^\alpha \right \} \quad .
\end{align}
The goal is now to expand the second term, $ \Tr \, \log \left \{ \bm{1} - \tilde{\bm{G}} \left[ \bm{J}^\alpha + \bm{\varphi}^\alpha \right] \bm{\gamma}^\alpha \right \}$, around the HB field configuration $\varphi_{\text{MF},r}$ that minimizes the action. 
Doing so, we can now collect all the relevant terms into the RPA action, given by
\begin{align}
& S_\text{RPA}[J,\varphi] = \int dx \, dx' \, \varphi^\alpha(x) \, [D^{-1}]^{\alpha \beta}(x,x') \, \varphi^\beta(x') \nonumber \\
& + \int dx \, dx' \,  \varphi^\alpha(x) \, \tilde{\chi}_0^{\alpha \beta}(x,x') \, J^\beta(x')  \nonumber \\
& + \int dx \, dx' \,  J^\alpha(x) \, \tilde{\chi}_0^{\alpha \beta}(x,x') \, \varphi^\beta(x')  \nonumber \\
& + \int dx \, dx' \, J^\alpha(x) \, \tilde{\chi}_0^{\alpha \beta}(x,x') \, J^\beta(x')  \quad ,
\label{S_RPA_1}
\end{align}
where we introduced $\tilde{\chi}_0$, a corrected version of the non-interacting susceptibility that takes into account the mean-field solution and is given by
\begin{equation}
\tilde{\chi}_0^{\alpha \beta}(x,x') =  \frac{i}{2} \Tr \left[ \bm{\gamma}^\alpha \, \tilde{\bm{G}}(x,x') \, \bm{\gamma}^\beta \, \tilde{\bm{G}}(x',x) \right] \quad .
\end{equation}
We also defined the renormalized inverse propagator of the HB fields
\begin{equation}
\bm{D}^{-1}(x,x') = U^{-1}_{rr'} \, \bm{\sigma}_x \, \delta(t-t') + \tilde{\bm{\chi}}_0(x,x') \quad .
\end{equation}
As can be seen the RPA action is quadratic in the HB fields, which can thus be integrated out, resulting in 
\begin{equation}
\log Z_\text{RPA}[J] = i \, \bm{J} \circ \tilde{\bm{\chi}}_0 \circ \bm{J}  - i \, \bm{J} \circ \tilde{\bm{\chi}}_0 \circ \bm{D} \circ \tilde{\bm{\chi}}_0 \circ \bm{J} \quad ,
\end{equation}
where the symbol $\circ$ stands for a convolution over space, time and the causal structure of the matrices. The RPA susceptibility is defined as
\begin{align}
\chi^{\alpha \beta}_\text{RPA} (x,x') = -\frac{i}{2} \left. \frac{ \delta^2 \, \log Z_\text{RPA}[J] }{ \delta J^\beta(x') \, \delta J^\alpha(x) } \right|_{J=0} \quad .
\end{align}
Performing the derivative and some manipulations finally leads to
\begin{equation}
\bm{\chi}_\text{RPA} = \tilde{\bm{\chi}}_0 \circ \left( \bm{1} + \bm{U} \, \bm{\sigma}_x \circ \tilde{\bm{\chi}}_0 \right)^{-1} \quad .
\label{P_RPA}
\end{equation}
The retarded component in frequency space is given by Eq.~(\textcolor{blue}{9}) of the main text.

\section{Frequency Integrals \label{Frequency Integrals}}

The calculation of the quantities analysed in the main text requires an integration over frequency space, see eqs.~(\textcolor{blue}{7}), (\textcolor{blue}{10}) and (\textcolor{blue}{11}) from the main text. For completeness sake, in this section we go over the details of this calculation.

In the main text we introduced the non-hermitian Hamiltonian $\bm{K}$, in terms of which the Green's functions are written. This matrix has left and right eigenvectors, $\ket{\alpha}$ and $\ket{\tilde{\alpha}}$ respectively, that obey the following properties
\begin{align}
& \bm{K} \ket{\alpha} = \epsilon_\alpha \ket{\alpha} \quad , \nonumber \\
& \bra{\tilde{\alpha}} \bm{K} = \bra{\tilde{\alpha}} \epsilon_\alpha \quad , \nonumber \\
& \bm{K} = \sum_\alpha \ket{\alpha} \, \epsilon_\alpha \, \bra{\tilde{\alpha}} \quad , \nonumber \\
& \langle \tilde{\alpha} | \alpha ' \rangle = \delta_{\alpha, \alpha'} \quad ,
\nonumber \\
& \sum_\alpha \ket{\alpha}\bra{\tilde{\alpha}} = \sum_\alpha \ket{\tilde{\alpha}} \bra{\alpha} = \bm{1} \quad ,
\end{align}
where $\epsilon_\alpha$ are the eigenvalues. Using these properties we can obtain an expression for the Green's functions in terms of the eigenvalues and eigenvectors 
\begin{align}
& \bm{G}^R(\omega) = \frac{1}{\omega - \bm{K}} = \sum_\alpha \ket{\alpha} \, \frac{1}{\omega - \epsilon_\alpha} \, \bra{\tilde{\alpha}} \quad , \nonumber \\
& \bm{G}^A(\omega) = \frac{1}{\omega - \bm{K}^\dagger } = \sum_\alpha \ket{\tilde{\alpha}} \, \frac{1}{\omega - \epsilon^*_\alpha} \, \bra{\alpha} \quad , \nonumber \\
& \bm{G}^K(\omega) = \bm{G}^R(\omega) \, \bm{\Sigma}^K(\omega) \, \bm{G}^A(\omega) \nonumber \\
&= -i \sum_{\alpha,\alpha',l} \text{sign} (\omega - \mu_l) \, \ket{\alpha} \, \frac{ \bra{\tilde{\alpha}} \bm{\Gamma}_l \ket{\tilde{\alpha'}} }{(\omega - \epsilon_\alpha) (\omega - \epsilon^*_{\alpha'}) } \, \bra{\alpha'} \quad .
\end{align}

From the covariance matrix, see Eq.~(\textcolor{blue}{7}) of the main text, we can obtain the electron density and the current. Using the relations above we can rewrite this equation according with 
\begin{equation}
\bm{\rho} = \frac{1}{2} \left[ \bm{1} - \sum_{\alpha,\alpha',l} I_l(\epsilon_\alpha, \epsilon^*_{\alpha'}) \, \ket{\alpha} \bra{\tilde{\alpha}} \bm{\Gamma}_l \ket{\tilde{\alpha}'} \bra{\alpha'}   \right]^T \quad ,
\end{equation}
where $I_l(\epsilon_\alpha, \epsilon^*_{\alpha'})$ are the frequency integrals given by
\begin{align}
& I_l(\epsilon_\alpha, \epsilon^*_{\alpha'}) = \int_{-\infty}^{+\infty} \frac{d \omega}{2\pi} \frac{ \text{sign} ( \omega - \mu_l ) }{ (\omega - \epsilon_\alpha) (\omega - \epsilon^*_{\alpha'} ) } \nonumber \\
&= \frac{1}{2\pi} \frac{ i\pi + \log[\epsilon^*_{\alpha'} - \mu_l] + \log[\mu_l - \epsilon^*_{\alpha'}] - 2 \log[\mu_l - \epsilon_{\alpha}] }{\epsilon_\alpha - \epsilon^*_{\alpha'}} .
\end{align}

A similar procedure can be used for the retarded and Keldysh susceptibilities.

\section{Logarithmic divergence of $\chi^R_0(0,q_\text{max})$}

%%%%%%%%%%%%%%%%%%%%%%% Logarithmic %%%%%%%%%%%%%%%%%%%%%%%%%%
\begin{figure}[t!]
\centering
\includegraphics[width= 0.45 \textwidth]{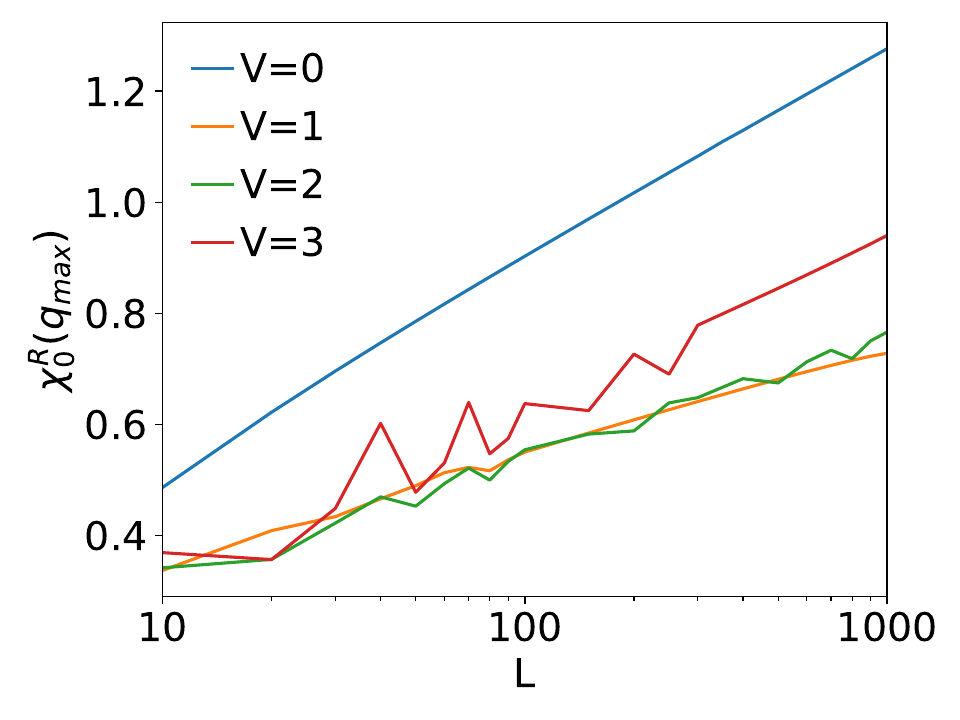} 
\caption{Real part of the static non-interacting susceptibility $\text{Re}\left[ \chi_0(\omega=0,q_\text{max}) \right]$ as a function of system size for different values of $V$. The x-axis is displayed in a logarithmic scale, showcasing the logarithmic divergence of the peaks of the susceptibility.}
\label{Logarithmic}
\end{figure}
%%%%%%%%%%%%%%%%%%%%%%%%%%%%%%%%%%%%%%%%%%%%%%%%%%%%%%%%%%

In Fig.~\textcolor{blue}{2}-(b) of the main text we show the real part of the static non-interacting susceptibility $\text{Re}\left[ \chi_0(\omega=0,q) \right]$ as a function of $q$ for different values of the bias $V$. The peaks in the susceptibility occur for $q_\text{max}(V) = \pi \pm 2\arcsin(V/4t)$. It was stated in the main text that the peaks of the susceptibility diverge logarithmically with system size. Figure~\ref{Logarithmic} provides evidence supporting that claim.

Under a RPA analysis the divergence of the bare susceptibility is connected to the appearance of instabilities already at infinitesimally small values of $U$ in the thermodynamic limit. These correspond to the onset of the $\text{CDW}_q$ phase.

\section{Typical Configurations}

\begin{figure*}
\centering
\includegraphics[width= 1.0 \textwidth]{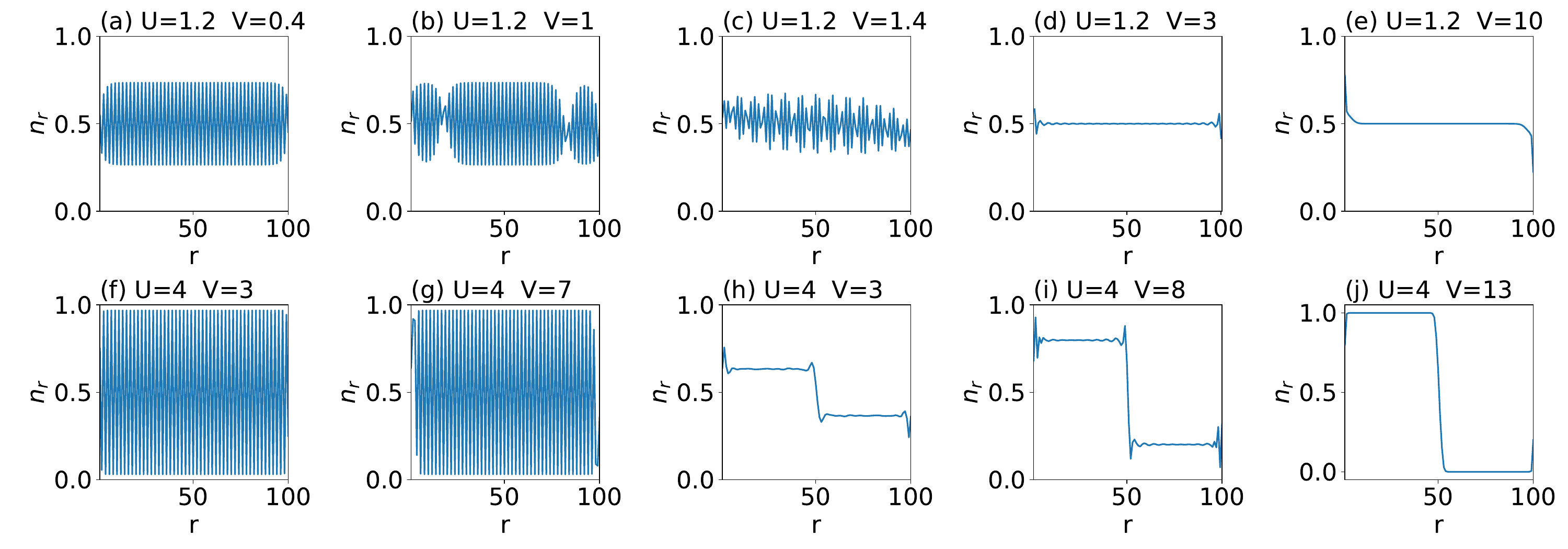} 
\caption{ Charge density profiles $n_r$ of the different phases. (a) and (f) correspond to the  $\text{CDW}_\pi$ at low and large coupling respectively. (b) and (g) also to the  $\text{CDW}_\pi$, but with boundary defects. (c) to the  $\text{CDW}_q$, (d) to C, (e) to MK-C, (h) to CDS-C, (i) to CDS-I and finally (j) to MK-I. (a)-(e) were obtained for $U=1.2$ and (f)-(j) for $U=4$. }
\label{dens_profiles}
\end{figure*}

Here, we provide charge density profiles $n_r$ characteristic of each of the phases shown in Fig.~\textcolor{blue}{1}-(b) of the main text.
Figs.~\ref{dens_profiles}-(a) to (e) do so for the weak coupling regimes at $U=1.2$ and Figs.~\ref{dens_profiles}-(f) to (j) are for strong coupling, {\itshape i.e.}, at $U=4$.

At weak coupling Fig.~\ref{dens_profiles}-(a) shows $n_r$ in the $\text{CDW}_\pi$ phase, (b) refers to the  $\text{CDW}_\pi$ as well, but after boundary defects have set in, (c) corresponds to the  $\text{CDW}_q$ phase, (d) to the C phase and finally (e) to the MK-C. The small oscillations in (d) vanish with increasing system size. This is how the  $\text{CDW}_q-$C transition line was determined, as stated in the main text.

At strong coupling, we have in (f) and (g) configurations referring to the  $\text{CDW}_\pi$ with and without boundary defects respectively, (h) to the CDS-C, (i) to the CDS-I and (j) to the MK-I. (h) was obtained for  backwards evolution. Only in this case does the CDS-C phase occur in the phase $U-V$ phase diagram.

%%%%%%%%%%%%%%%%%%%%%%% Figure X %%%%%%%%%%%%%%%%%%%%%%%%%%
%%%%%%%%%%%%%%%%%%%%%%%%%%%%%%%%%%%%%%%%%%%%%%%%%%%%%%%%%%

\section{Boundary Truncation}

\begin{figure}[t!]
\centering
\includegraphics[width= 0.48 \textwidth]{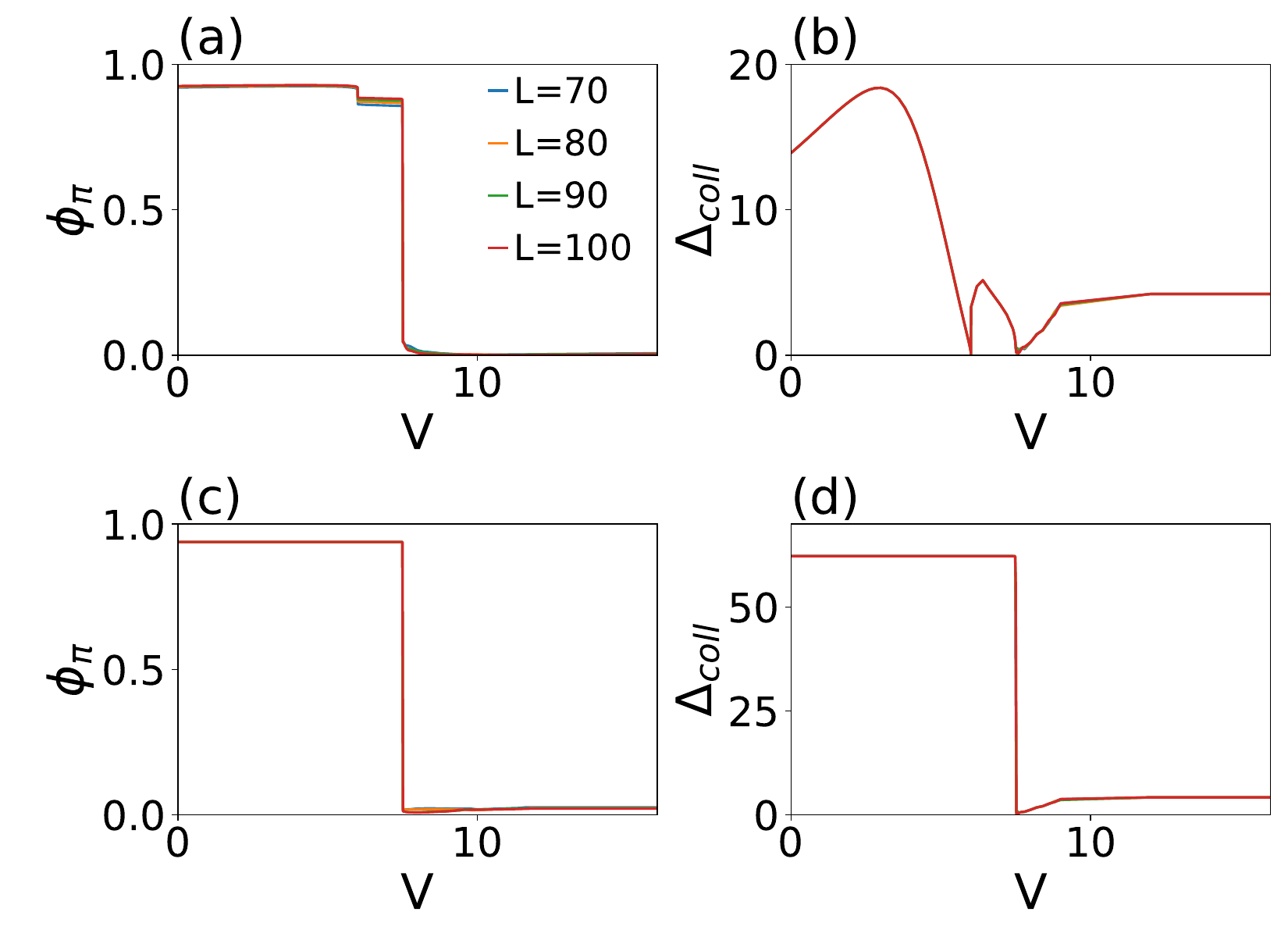} 
\caption{ Effect of truncating the system in some physical quantities. (a) and (b) show the order parameter $\phi_\pi$ and the collective excitation gap $\Delta_\text{coll}$ without truncation. (c) and (d) show the same quantities with truncation. The results were obtained for $U=4$. }
\label{Truncation}
\end{figure}

As stated in the main text, the calculation of some of the quantities that were analysed involve the truncation of the charge density profile $n_r$. As seen in Fig.~\ref{dens_profiles}, some of the phases contain boundary defects, that affect the value of these quantities. To illustrate this point, we show in Fig.~\ref{Truncation}, for the large bias transition, the order parameter $\phi_\pi$ and the collective excitation gap $\Delta_\text{coll}$, calculated in terms of the retarded RPA susceptibility matrix $\bm{\chi}^R_\text{RPA}(\omega=0)$. (a) and (b) show these quantities without truncation and (c) and (d) with. We see in (a) that the onset of boundary defects has a noticeable effect on the order parameter. This, however, subsides with increasing system size. Likewise, the collective excitation gap is also affected by the boundary defects. Upon truncation both these signals disappear.

%%%%%%%%%%%%%%%%%%%%%%% Truncation %%%%%%%%%%%%%%%%%%%%%%%%%%
%%%%%%%%%%%%%%%%%%%%%%%%%%%%%%%%%%%%%%%%%%%%%%%%%%%%%%%%%%

\section{Critical Point of MK-C Phase}

It was stated in the main text that the MK-C phase is separated by a phase transition line, signalled by the vanishing of the collective excitation gap $\Delta_\text{coll}$ for an interaction coupling of around $U \approx 0.9$. This line ends in a critical point at $V^* \approx 29$. In Fig.~\textcolor{blue}{4}-(c) of the main text we show the $\Delta_\text{coll}$ as a function of $U$ for two different values of the bias $V=16$ and $V=100$, respectively bellow and above the critical point. For $V=16$ $\Delta_\text{coll}$ vanishes at $U \approx 0.9$ and for $V=100$ the corresponding minimum of $\Delta_\text{coll}$ is still finite.  

\begin{figure}[t!]
\centering
\includegraphics[width= 0.48 \textwidth]{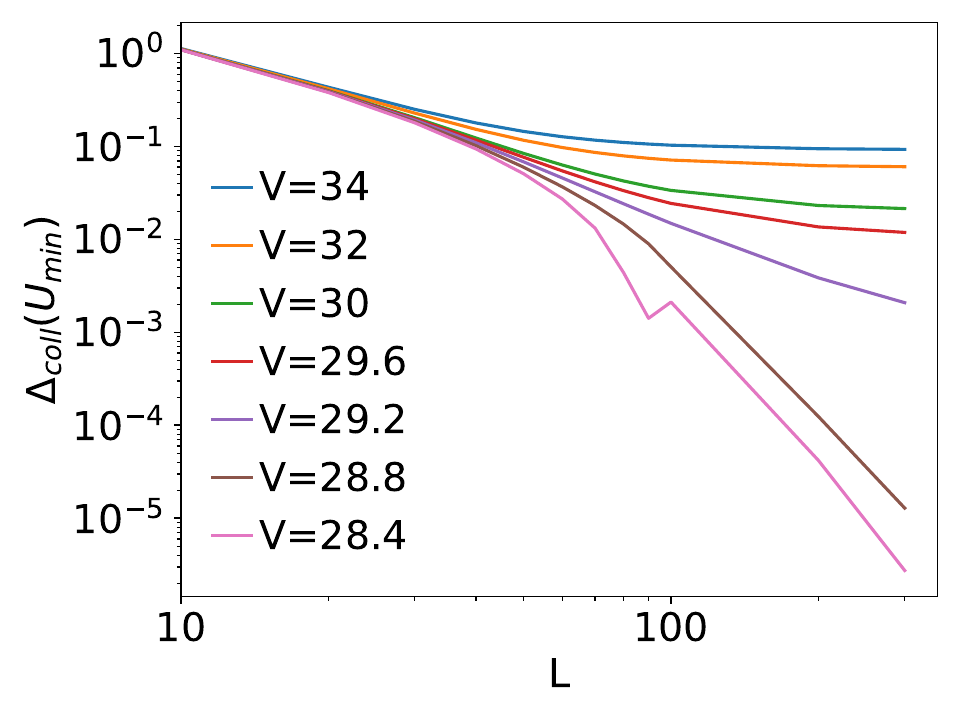} 
\caption{Minimum of the collective excitation gap $\Delta_\text{coll}(U_c)$ as a function of system size $L$ for different values of the bias $V$. Both axes are shown in a logarithmic scale.}
\label{Critical_Point}
\end{figure}

Here we show how the critical value of $V^* \approx 29$, reported in the main text, was obtained. In Fig.~\ref{Critical_Point} we show the collective excitation gap $\Delta_\text{coll}$, evaluated for the $U$ that minimizes this function, as a function of system size for different values of the bias. For values of $V>V^*$, we note that $\Delta_\text{coll}$ converges with $L$ to a finite value and for $V<V^*$ it converges to zero.

%%%%%%%%%%%%%%%%%%%%%%% Critical Point %%%%%%%%%%%%%%%%%%%%%%%%%%
%%%%%%%%%%%%%%%%%%%%%%%%%%%%%%%%%%%%%%%%%%%%%%%%%%%%%%%%%%

\section{Simplified Mean-Field Model}

In the main text, when analysing the strong coupling regimes, we introduced a simplified version of the mean-field procedure. In this section, we provide further details on this method and discuss some exact results for the determination of the transition lines.

The simplified model works by reducing the mean-field procedure to a single parameter $n$ that gives the occupation of the sites on the left side of the system. On the right side the occupation is thus $1-n$. The center of the energy bands on the left and right side of the system is then taken to be $\tilde{\varepsilon}_{L/R} = \pm U (2n-1)$. As stated before, their bandwidth is $W=4$.

\begin{figure}[t!]
\centering
\includegraphics[width= 0.48 \textwidth]{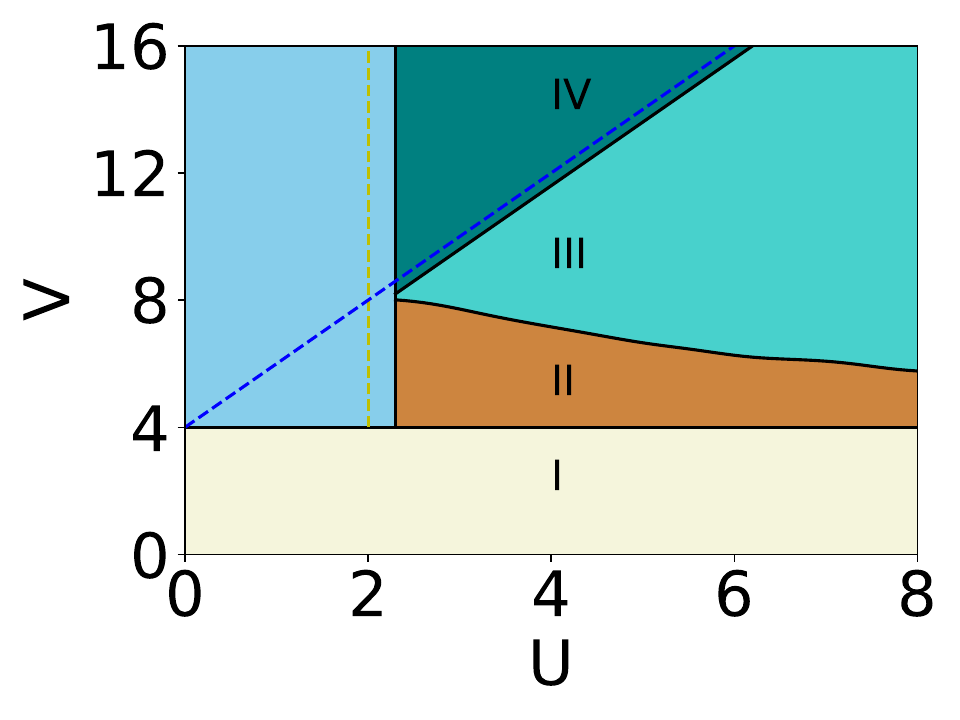} 
\caption{ Phase diagram of the simplified mean-field model for $L=100$. The analytic prediction for some of the phase transitions is added in dashed blue and yellow lines.}
\label{Dummy}
\end{figure}

It is now possible to establish analytically the transition lines that enclose the IV regime. The III--IV transition happens when the bias reaches the end of the bands, {\itshape i.e.}, when $V/2 = U(2n-1) + 2$. Given that at this transition the occupation saturates ($n=1$) we have $V = 2U + 4$ as the transition line. In Fig.~\ref{Dummy} we replicate the phase diagram of the simplified model and added this line in dashed blue, where a relatively good agreement with the numeric result can be seen. 

Starting in the IV phase and lowering $U$ the system transitions into a conducting phase that is still phase separated. It was ignored in the main text because the simplified model is only expected to replicate the strong coupling results of the full model. The transition line occurs when the top of the right band touches the bottom of the one on the left $U(2n-1)-2=-U(2n-1)+2$. Using that the occupation is saturated yields $U=2$, which is illustrated in Fig.~\ref{Dummy} in dashed yellow. The slight discrepancy between the analytic prediction and the numeric result is likely due to finite-size effects.

%%%%%%%%%%%%%%%%%%%%%%% Dummy %%%%%%%%%%%%%%%%%%%%%%%%%%
%%%%%%%%%%%%%%%%%%%%%%%%%%%%%%%%%%%%%%%%%%%%%%%%%%%%%%%%%%

\section{Markov Limit}

It was stated in the main text that the large bias limit $V\to \infty$ corresponds to the Markovian regime. In this Section we provide further details to bring out this connection. In Ref.~\cite{Sieberer_2016}, a Keldysh action is built from the Lindblad master equation, allowing both formulations to be treated on equal footing.

Let us consider the model given by the following Lindblad equation
\begin{equation}
\partial_t \, \rho = -i [H,\rho] + \sum_l \left( W^\dagger_l \, \rho \, W_l - \frac{1}{2} \{ W^\dagger_l \, W_l , \rho \} \right) ,
\label{Lindblad}
\end{equation}
where $[]$ and $\{ \}$ correspond to commutators and anti-commutators respectively, $\rho$ is the density matrix and the Hamiltonian $H$ is the same as in Eq.~(\textcolor{blue}{1}) of the main text. The sum goes over the left and right reservoir with labels $l=1,L$ respectively and the jump operators $W_l$ are given by
\begin{equation}
W_1 = \sqrt{\eta} \, c^\dagger_1 \quad \text{and} \quad W_L = \sqrt{\eta} \, c_L \quad ,
\end{equation}
where $\eta$ is the particle injection/removal rate. This corresponds to the case where particles are only injected into the system from the left and removed from the right at the same rate. This model has received substantial attention and has an exact solution in terms of a Matrix Product State (MPS) parametrization~\cite{Prosen3,Prosen4}. 

Employing the same prescription as in Ref.~\cite{Sieberer_2016}, we obtain the following Keldysh action for this model
\begin{align}
S &=  \int d\omega
\begin{pmatrix}
\bm{\bar\Psi}_1  \\
\bm{\bar\Psi}_2  \\
\end{pmatrix}^T
\begin{pmatrix}
\bm{P}^R(\omega) & \bm{P}^K(\omega)  \\
0 & \bm{P}^A(\omega) \\
\end{pmatrix}
\begin{pmatrix}
\bm{\Psi}_1  \\
\bm{\Psi}_2  \\
\end{pmatrix} + S_\text{int}  ,
\end{align}
where $\bm{\Psi}_1$ and $\bm{\Psi}_2$ are Keldysh rotated fermion fields, as in Eq.~\eqref{rotation}. The interaction action $S_\text{int}$ has the same form as in our model, see Eq.~\eqref{interacting_action}, and since it does not depend on the bias it takes no part in the $V\to \infty$ limit. The field's inverse propagators have the following expressions
\begin{align}
& \bm{P}^R(\omega) = \left[\bm{P}^A(\omega)\right]^\dagger = \omega - \bm{H} + i \, \frac{ \bm{\Gamma}_\text{in} + \bm{\Gamma}_\text{out} }{2} \nonumber \\
& \bm{P}^K(\omega) = i \,  \bm{\Gamma}_\text{out} -  i \,  \bm{\Gamma}_\text{in} \quad ,
\end{align}
where the matrices $\left[\bm{\Gamma}_\text{in}\right]_{r,r'} = \eta \, \delta_{r,r'} \, \delta_{r,1}$ and $\left[\bm{\Gamma}_\text{out}\right]_{r,r'} = \eta \, \delta_{r,r'} \, \delta_{r,L}$ are associated with the left and right reservoirs respectively.

We can see already that the retarded and advanced components have the same form as in our model, see eqs.~\eqref{Inverse_Green} and \eqref{self-retarded-freq}, so long as we set the following equality between the parameters of both models $\eta=2\pi \, \rho \, J^2$. $\rho$ is the reservoir's density of states and $J$ the reservoir-system hopping; and as stated in the main text we set $J=\rho=1$ for both reservoirs. We deal now with the Keldysh component. For zero temperature, the hyperbolic tangent in Eq.~\eqref{self-Keldysh-freq} reduces to a sign function and upon the limit $\mu_1 = -\mu_l = V/2 \to \infty$ we obtain the same expression as in $\bm{P}^K(\omega)$. We have thus shown that the $V\to \infty$ limit corresponds to the Markovian regime.

%%%%%%%%%%%%%%%%%%%%%%% Prosen %%%%%%%%%%%%%%%%%%%%%%%%%%
\begin{figure}
\centering
\includegraphics[width= 0.5 \textwidth]{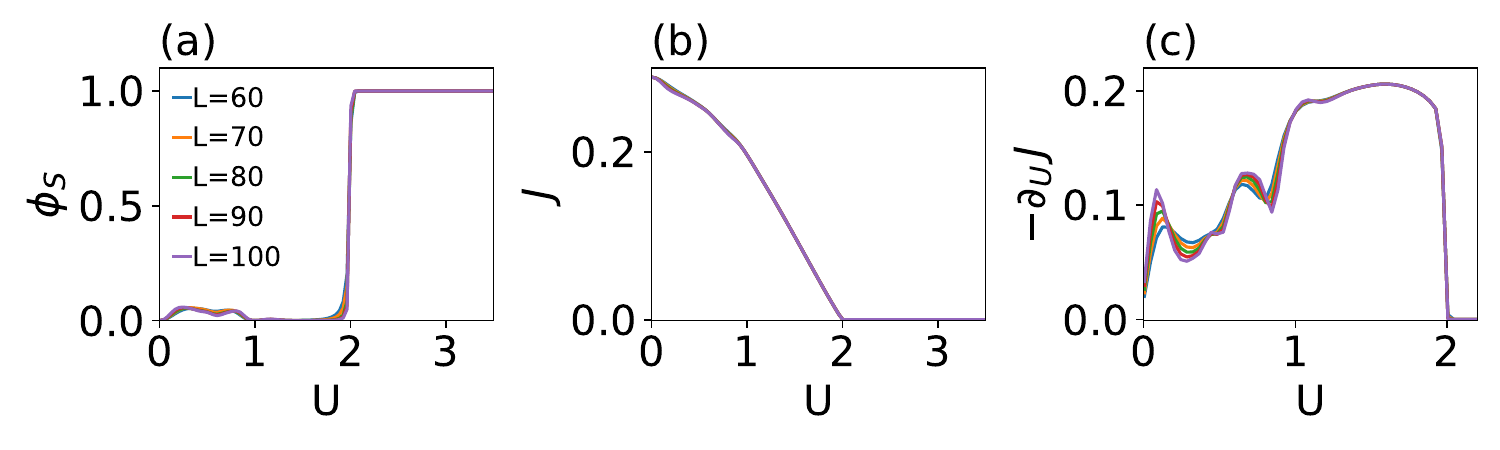} 
\caption{ Results using Prosen's exact solution. (a) shows the charge separated order parameter $\phi_S$ as a function of $U$ for different system sizes, (b) the current $J$ and (c) the negative differential conductivity $-\partial_U \, J$.}
\label{Prosen}
\end{figure}
%%%%%%%%%%%%%%%%%%%%%%%%%%%%%%%%%%%%%%%%%%%%%%%%%%%%%%%%%%

%%%%%%%%%%%%%%%%%%%%%%% Markov %%%%%%%%%%%%%%%%%%%%%%%%%%
\begin{figure*}
\centering
\includegraphics[width= 0.8 \textwidth]{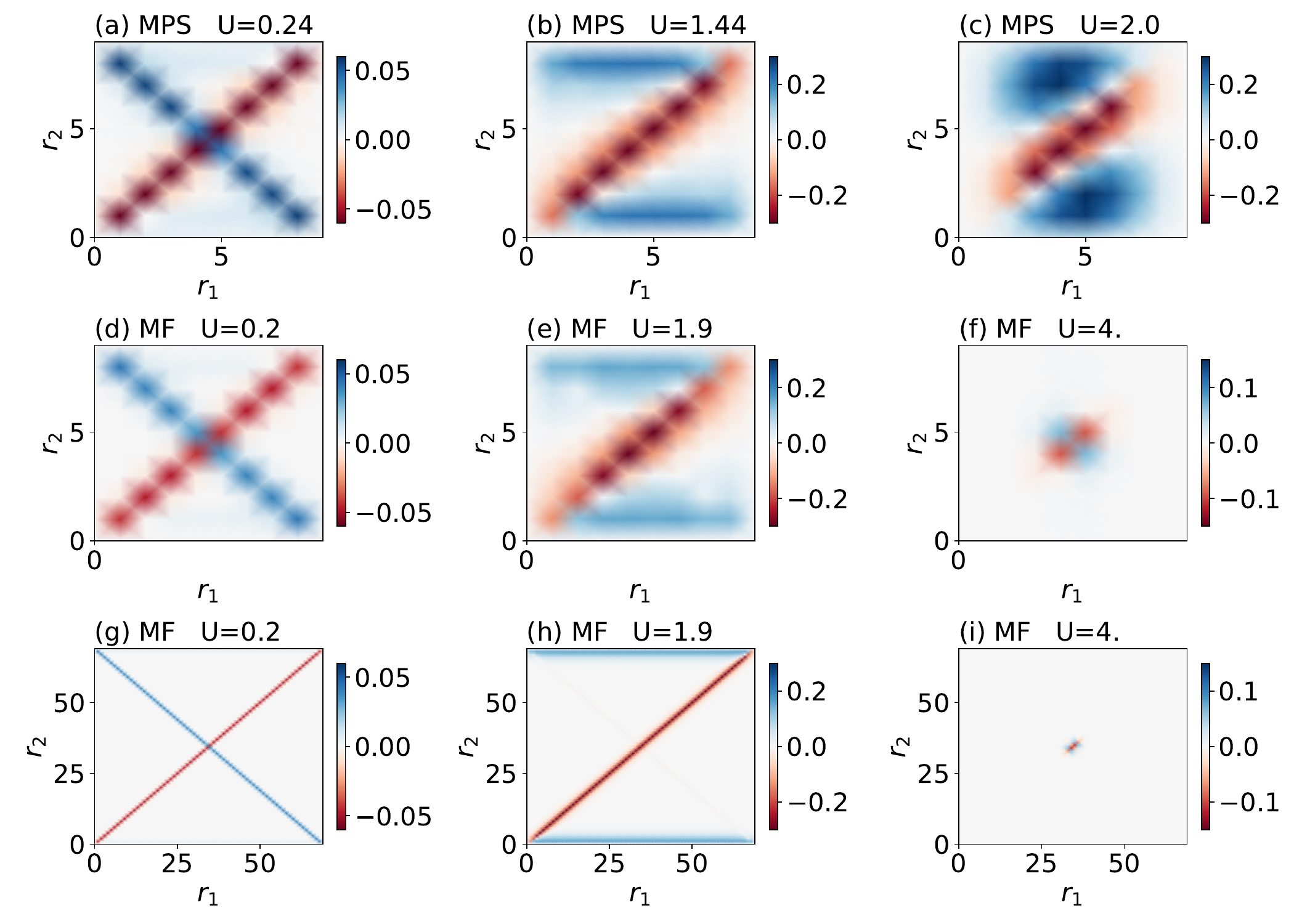} 
\caption{ (a) to (c) retarded static susceptibility of Markov model computed using MPS methods for different values of $U$. (d) to (i) retarded RPA static susceptibility computed using mean-field for $V=100$. (a) to (f) were computed for $L=10$ and (g) to (i) for $L=70$.}
\label{Markov}
\end{figure*}
%%%%%%%%%%%%%%%%%%%%%%%%%%%%%%%%%%%%%%%%%%%%%%%%%%%%%%%%%%

By way of a Jordan–Wigner transformation, the Markovian fermionic model in Eq.~\eqref{Lindblad} coincides with the spin model in refs.~\cite{Prosen3,Prosen4}, where an exact solution is provided. In Fig.~\ref{Prosen} we show some quantities computed using this exact solution: in (a) the order parameter $\phi_S$, in (b) the current $J$ and in (c) the negative differential conductivity $-\partial_U \, J$. We can now compare these with the large bias results in Fig.~\textcolor{blue}{4} of the main text. The MK-C--MK-I transition happens for $U_c=2$ in the exact model and for $U_c\approx 2.56$ in the mean-field description. These kind of shifts are typical in mean-field solutions; the two phases are however qualitatively well captured by it. We can see  pronounced finite size effects for $U<1$ that are not present in mean-field. However, for small values of $U$ there are a number of singularities in the collective excitation gap $\Delta_\text{coll}$, see Figs.~\textcolor{blue}{4}-(c), that are possibly a remnant of the structure of the exact model.

Additionally, in Fig.~\ref{Markov} we look at the static retarded susceptibility in real space $\chi(\omega=0)$ of the full model across the MK-C--MK-I transition, see (a) to (c), and compare it with the mean-field RPA susceptibility $\chi^R_\text{RPA}(\omega=0)$, see (d) to (i). The susceptibility is not accessible using Prosen's exact solution, presented in refs.~\cite{Prosen3,Prosen4}, however it can be approximated using a MPS method adapted for non-equilibrium systems. We were also limited to small system sizes, given the complexity and stability of the calculation. We followed the formalism described in the supplemental material of Ref.~\cite{M.Oliveira2023}.

Starting the comparison at small system size $L=10$ and for small values of $U$, both (a) and (d) show a symmetric retarded response, where a perturbation applied to site $r$ causes an anti-symmetric response on site $L-r$. This effect subsides for larger values of $U$, but that are still in the MK-C phase, see (b) and (e). This difference in the response could be concomitant with the difference in finite size effects that are present in the exact model for small interaction strength and then subside with increasing $U$. In the MK-I phase the response seems to be confined to the middle of the system, where the domain wall is located, see (f). We could not use the MPS method to probe inside the MK-I phase; this is due to the exponentially slow convergence rate of the method~\cite{Benenti1}. However at the transition point $U=2$ we already see the response starting to localize around the middle of the system, see (c). The same features are still present in the mean-field model for larger system sizes, see (g) to (i) for $L=70$.

The qualitative agreement between the results obtained for the exact solution and the mean-field method strengthens the legitimacy of the application of the later in this study.

\end{document}